\documentclass[]{article}
\usepackage{mathtools}
\usepackage{graphicx}
\usepackage{xcolor}
\usepackage{upgreek}
\usepackage{float}
\usepackage{txfonts}
\usepackage{longtable}
\usepackage{caption}

\usepackage[a4paper, left=2cm, right=2cm, top=2cm, bottom=2cm, includeheadfoot]{geometry}

\usepackage{multirow}
\usepackage{multicol}

\usepackage[comma,super,sort&compress]{natbib}
\usepackage[colorlinks = true,
            linkcolor = blue,
            urlcolor  = blue,
            citecolor = blue,
            anchorcolor = blue]{hyperref}

\newcommand{\apj}{Astrophys. J.} %{ApJ}    
\newcommand{\apjs}{Astrophys. J. Suppl. Ser.} %{ApJS}       
                                 
\newcommand{\mnras}{Mon. Not. R. Astron. Soc.} %{MNRAS}     
\newcommand{\nat}{Nature}

\newcommand{\aap}{Astron. Astrophys.} %{A\&A}            
                                
\newcommand{\aapr}{Astron. \& Astrophys.\ Rev.} %{A\&AR}                                
\newcommand{\apjl}{Astrophys. J. Lett.} %{ApJL}          
                                
\newcommand{\pasj}{PASJ}                                 
\newcommand{\pasp}{Publ. Astron. Soc. Pac.} %{PASP}       
\newcommand{\pasa}{Pubs. Astron. Soc. Australia}

\newcommand{\rone}{FRB~20121102A} % for FRB

\makeatother
\usepackage[switch]{lineno} % for the line numbers

\title{}

\newcommand{\MPS}{\href{https:/orcid.org/0000-0001-6170-2282}{\textcolor{blue!50!black}{M.~P.~Snelders}}}
\newcommand{\KN}{\href{https:/orcid.org/0000-0003-0510-0740}{\textcolor{blue!50!black}{K.~Nimmo}}}
\newcommand{\JWTH}{\href{https:/orcid.org/0000-0003-2317-1446}{\textcolor{blue!50!black}{J.~W.~T.~Hessels}}}
\newcommand{\ZB}{\href{https:/orcid.org/0009-0005-3353-9592}{\textcolor{blue!50!black}{Z.~Bensellam}}}
\newcommand{\LPZ}{\href{https:/orcid.org/0000-0002-4940-5610}{\textcolor{blue!50!black}{L.~P.~Zwaan}}}
\newcommand{\PC}{\href{https:/orcid.org/0000-0002-3426-7606}{\textcolor{blue!50!black}{P.~Chawla}}}
\newcommand{\OSOB}{\href{https:/orcid.org/0000-0001-9381-8466}{\textcolor{blue!50!black}{O.~S.~Ould-Boukattine}}}
\newcommand{\FK}{\href{https:/orcid.org/0000-0001-6664-8668}{\textcolor{blue!50!black}{F.~Kirsten}}}
\newcommand{\JTF}{\href{https:/orcid.org/0000-0001-9855-5781}{\textcolor{blue!50!black}{J.~T.~Faber}}}
\newcommand{\VG}{\href{https:/orcid.org/0000-0002-8604-106X}{\textcolor{blue!50!black}{V.~Gajjar}}}

\newcommand{\ASTRON}{ASTRON, Netherlands Institute for Radio Astronomy, Oude Hoogeveensedijk 4, 7991 PD Dwingeloo, The Netherlands}
\newcommand{\CAHILL}{Cahill Center for Astronomy and Astrophysics, MC 249-17 California Institute of Technology, Pasadena CA 91125, USA}
\newcommand{\CUT}{Department of Space, Earth and Environment, Chalmers University of Technology, Onsala Space Observatory, 439 92, Onsala, Sweden}
\newcommand{\MITK}{MIT Kavli Institute for Astrophysics and Space Research, Massachusetts Institute of Technology, 77 Massachusetts Ave, Cambridge, MA 02139, USA}
\newcommand{\UVA}{Anton Pannekoek Institute for Astronomy, University of Amsterdam, Science Park 904, 1098 XH Amsterdam, The Netherlands}
\newcommand{\BLBERKELEY}{Breakthrough Listen, University of California, Berkeley, 501 Campbell Hall 3411, Berkeley, CA, 94720, USA}
\newcommand{\SETI}{SETI Institute, 339 Bernardo Ave Suite 200 Mountain View, CA 94043, USA}

\author{}

\begin{document}

% \noindent{\LARGE \textbf{Microsecond-duration bursts from FRB~20121102A}} \\
\noindent{\LARGE \textbf{Detection of ultra-fast radio bursts from FRB~20121102A}} \\

\hfill \\

\noindent{\Large \MPS$^{\lozenge, 1, 2}$, \KN$^{3}$, \JWTH$^{2, 1}$, \ZB$^{2}$, \LPZ$^{2}$, \PC$^{2}$, \OSOB$^{1,2}$, \FK$^{4}$, \JTF$^{5}$ and \VG$^{6, 7}$} \\
\hfill \\

{\large\noindent$^{\lozenge}$ \url{snelders@astron.nl} \vspace{0.15cm} \\
\noindent$^{1}$ \textit{\ASTRON} \vspace{0.15cm} \\
\noindent$^{2}$ \textit{\UVA} \vspace{0.15cm} \\
\noindent$^{3}$ \textit{\MITK} \vspace{0.15cm} \\
\noindent$^{4}$ \textit{\CUT} \vspace{0.15cm} \\
\noindent$^{5}$ \textit{\CAHILL} \vspace{0.15cm} \\
\noindent$^{6}$ \textit{\SETI} \vspace{0.15cm} \\
\noindent$^{7}$ \textit{\BLBERKELEY} \vspace{0.15cm} \\}

\begin{multicols}{2}
\section*{Abstract}\label{abstract}
\textbf{Fast radio bursts (FRBs) are extragalactic transients with typical durations of milliseconds. FRBs have been shown, however, to fluctuate on a wide range of timescales: some show sub-microsecond sub-bursts while others last up to a few seconds in total. Probing FRBs on a range of timescales is crucial for understanding their emission physics, how to detect them effectively, and how to maximize their utility as astrophysical probes. \rone\ is the first-known repeating FRB source. Here we show that \rone\ is able to produce isolated microsecond-duration bursts whose total durations are more than ten times shorter than all other known FRBs to date. The polarimetric properties of these micro-bursts resemble those of the longer-lasting bursts, suggesting a common emission mechanism producing FRBs spanning a factor of 1,000 in duration. Furthermore, this work shows that there exists a population of ultra-fast radio bursts that current wide-field FRB searches are missing due to insufficient time-resolution. These results indicate that FRBs occur more frequently and with greater diversity than initially thought. This could also influence our understanding of energy, wait time, and burst rate distributions.}

% Introduction
\section*{Introduction}\label{intro}
FRBs are broadly divided into two categories: repeating and apparently one-off FRBs (see \cite[Petroff~et~al.][]{petroff_2019_aarv, petroff_2022_aarv} for recent reviews). Although repeaters and one-offs typically exhibit different observational properties (e.g., repeating FRBs generally show narrower spectral and wider temporal extents than non-repeaters \cite{pleunis_2021_apj}), it remains unclear whether this is the result of a different source class, emission physics, propagation effects or beaming geometry\cite{conner_2020_mnras_497}.

While repeating sources represent only a few percent of all known FRBs\cite{chime_2023_apj}, they are important for understanding the phenomenon. The repeatability allows for very long baseline interferometry (VLBI) localisations to probe their local environments \cite{marcote_2017_apjl,marcote_2020_natur,kirsten_2022_natur,nimmo_2022_apjl}; long-term monitoring to explore how their properties vary with time (e.g., scattering timescale\cite{ocker_2023_mnras}, Faraday rotation measure, RM \cite{michilli_2018_natur,hilmarsson_2021_apjl} and dispersion measure, DM\cite{spitler_2014_apj, hessels_2019_apjl, josephy_2019_apjl, wang_2022_atel_15619}); mapping of their sometimes periodic activity rate \cite{chime_2020_natur_periodicactivity}; and also multi-epoch, multi-wavelength searches spanning radio wavelengths \cite{pleunis_2021_apjl,gajjar_2018_apj} to high energies \cite{magiccollaboration_2018_mnras, hiramatsu_2023_apjl}.

The first-discovered repeating FRB, \rone \cite{spitler_2016_natur}, is one of the best-studied sources. Not only did the repeating nature of \rone\ rule out cataclysmic models to explain all FRBs \cite{spitler_2016_natur}, but \rone\ was also the first FRB to be precisely localised to a host galaxy \cite{chatterjee_2017_natur}, confirming the extragalactic distances to FRBs implied by their high DMs \cite{lorimer_2007_sci}. \rone\ lives in the outskirts of a star-forming region in a dwarf galaxy at a luminosity distance of approximately 1\,Gpc \cite{chatterjee_2017_natur,tendulkar_2017_apjl,marcote_2017_apjl,bassa_2017_apjl}. The FRB emitter is associated with a compact persistent radio source (PRS), which shows that it may be in a dense nebula or in the near vicinity of a massive black hole \cite{marcote_2017_apjl}. This is further supported by the discovery that \rone\ is in an extreme and dynamic magneto-ionic environment, as shown by its exceptionally high and variable RM\cite{michilli_2018_natur,gajjar_2018_apj}. The RM is variable on a range of timescales\cite{michilli_2018_mnras}, from days to years, and has decreased from about $127\times10^{3}$\,rad\,m$^{-2}$ to about $31\times10^{3}$\,rad\,m$^{-2}$ (in the observer frame) over the span of $7$\,years\cite{plavin_2022_mnras,hilmarsson_2021_apjl,feng_2023_atel}.

While most surveys search for FRBs on timescales of milliseconds, in some cases the raw voltage data are saved, allowing for extremely detailed studies of FRBs probing the emission on much shorter timescales\cite{cho_2020_apjl}. The as-yet non-repeating FRB~20170827A was shown to exhibit a $0.5$\,ms burst envelope, with a narrow sub-component of duration roughly $30$\,$\upmu$s \cite{farah_2018_mnras}. Similarly, \rone\ occasionally shows comparably narrow temporal features within a broader envelope\cite{michilli_2018_natur}. More recently, the repeating sources FRB~20180916B and FRB~20200120E were shown to have temporal sub-structures with durations of microseconds\cite{nimmo_2021_natas} down to tens of nanoseconds\cite{nimmo_2022_natas}, respectively. Timescales shorter than tens of microseconds remain unexplored for \rone, despite micro and sub-microsecond burst structure having been observed in other repeating FRBs \cite{farah_2018_mnras, nimmo_2021_natas,nimmo_2022_natas}. To date, these short-timescale variations have only been observed within broader burst envelopes with durations of at least $0.1$\,ms or longer.

Bursts from \rone\ show a variety of different morphologies, from simple single Gaussian burst profiles, to complex drifting islands of emission, now known to be characteristic of repeating FRB morphology\cite{hessels_2019_apjl,pleunis_2021_apj}. \rone\ bursts also often appear to be narrowband (roughly 20\,\% fractional bandwidth), and have been seen to emit at frequencies between $600$\,MHz \cite{josephy_2019_apjl} and $8$\,GHz \cite{gajjar_2018_apj}. These $8$\,GHz detections with the Green Bank Telescope (GBT) \cite{gajjar_2018_apj} remain the highest-frequency FRB detections, to date. The $4$--$8$\,GHz GBT data presented in \citet{gajjar_2018_apj} were searched for bursts at a time-resolution of $350$\,$\upmu$s, discovering a total of $21$ bursts in the first hour of a $5$-hour observing block, with $18$ bursts occurring within the first $30$\,minutes. A re-analysis of this same dataset by \citet{zhang_2018_apj}, retaining the same temporal and spectral resolution but using a machine learning search technique (differing from the standard boxcar matched-filtering technique used in \citet{gajjar_2018_apj}), discovered an additional $72$ candidate bursts, with half of the bursts occurring within the first $30$\,minutes. 

In this paper, we present a re-analysis of the first $30$\,minutes of the \citet{gajjar_2018_apj} GBT dataset, searching for ultra-fast radio bursts on timescales of microseconds: a parameter space that the previous searches of this dataset, and current FRB searches in general, are insensitive to. 

\section*{Observations and burst search}
A $5$-hour observation of \rone\ was carried out on 26 August 2017 using the GBT with the Breakthrough Listen (BL) recording system\cite{macmahon_2018_pasp}. The BL system recorded from $3.9$\,GHz to $9.3$\,GHz, fully covering the sensitivity range of the GBT C-band receiver (Methods). Raw voltages of the dual-linear receiver were recorded and stored allowing for offline re-processing, coherent dedispersion, high-time-resolution searches and polarimetric analysis.

Here we have analyzed the voltage data of the first $30$\,minutes of the observation, totalling more than $32$\,terabyte (TB) in size (Methods). The voltage data were coherently dedispersed with a DM of $560.5$\,pc\,cm$^{-3}$ (from \citet{hessels_2019_apjl}) using \texttt{digifil}\cite{vanstraten_2011_pasa}, removing the intra-channel smearing but keeping the inter-channel smearing across the band. The resulting total-intensity filterbanks\cite{lorimer_2011_ascl} have a time-resolution of $341.\bar{3}$\,ns and $4.5$\,GHz of bandwidth at a centre frequency of $6126$\,MHz. 

To save a factor of greater than $10$ in computational costs, we downsample the filterbanks in time using \texttt{digifil} to create new filterbank data products with a time-resolution of approximately $2$, $33$ and $524$\,$\upmu$s, respectively. For a hypothetical burst with a duration of $341.\bar{3}$\,ns, this reduces the sensitivity by a factor of $\sqrt{6}$ compared to using the full available time resolution offered by the data (Methods). Since bursts from \rone\ often appear narrowband, we choose to do multiple subbanded searches: dividing the total bandwidth into subbands of widths $4.5$\,GHz, $1.5$\,GHz (the spectral width of the widest bursts in \citet{gajjar_2018_apj}), $750$\,MHz, $375$\,MHz, and $187.5$\,MHz (the spectral width of the bright patches of the bursts in \citet{gajjar_2018_apj}) before searching the data for FRBs.

We searched for bursts over a range of DMs centred around $560.5$\,pc\,cm$^{-3}$ using the \texttt{PRESTO}\cite{ransom_2001_phdt} suite of tools and very small step-sizes in trial DM (Methods). Microsecond-duration bursts require small step-sizes in DM because their recovered S/N decreases rapidly if the assumed DM is incorrect\cite{cordes_2003_apj}. At these high frequencies the data are relatively clean and thus we did not apply any radio frequency interference (RFI) masks to avoid the accidental masking of any bright and/or short-duration bursts. Candidate astrophysical signals were classified using the machine learning classifier \texttt{FETCH}\cite{agarwal_2020_mnras}. Candidates with a \texttt{FETCH}-assigned probability $p \geq 0.5$ of being astrophysical or temporal duration $\leq 500$\,$\upmu$s (regardless of the probability given by \texttt{FETCH}) were manually inspected.

\section*{Results}\label{results}
\begin{figure*}
    \centerline{\includegraphics[width=1.2\textwidth]{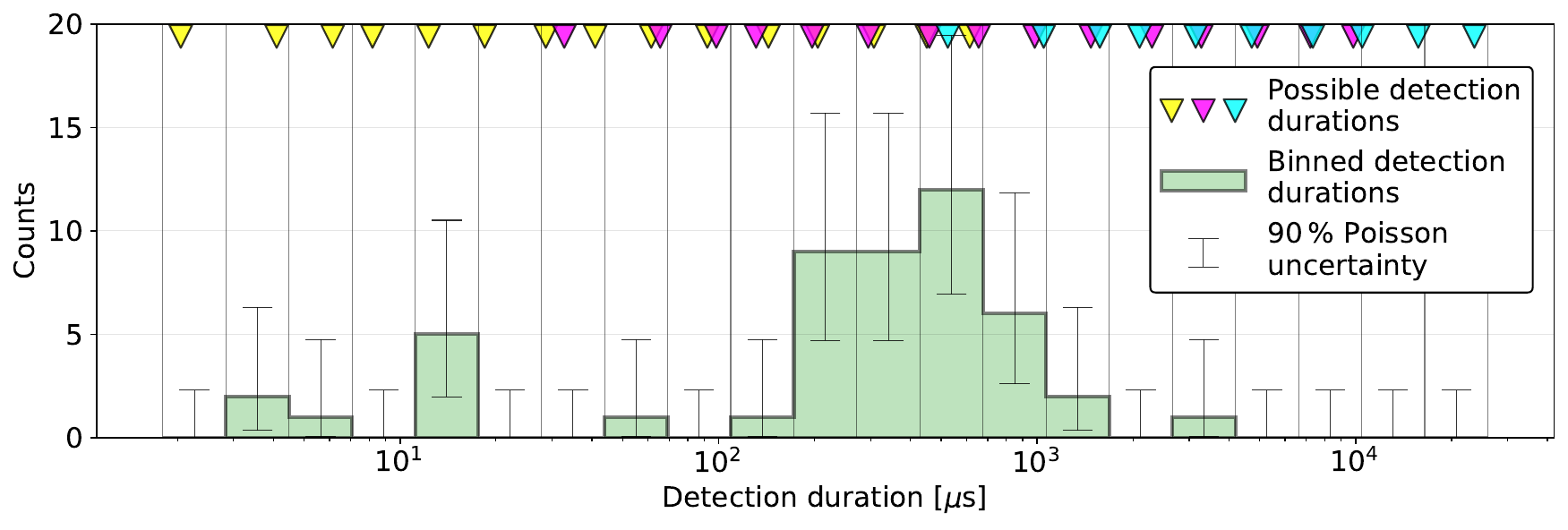}}
    \caption{\textbf{Histogram of the bursts durations, as they were found in the search.} The majority of the bursts have a typical duration of roughly $0.5$\,ms, but 8 bursts have extremely short durations of less than about $15$\,$\upmu$s, with some of them as short as roughly $4$\,$\upmu$s. The coloured triangles indicate the possible detection durations, i.e., they are the boxcar widths used in \texttt{PRESTO}'s \texttt{single\_pulse\_search.py} (Methods). The colours yellow, magenta and cyan represent boxcars where the underlying sampling time of the data was $2$, $33$ and $524$\,$\upmu$s, respectively. The black asymmetric error bars are centred on the number of counts per bin and represent the $90$\,\% Poisson uncertainty range of counts in each bin. Only one burst with a temporal width between about $20$\,$\upmu$s and $100$\,$\upmu$s was found and none of the bursts have a temporal width greater than about $4$\,ms, even though the search was sensitive to those timescales.}
    \label{fig:dur_hist}
\end{figure*}

We find 49 unique bursts within the first $30$\,minutes of the observation (Supplementary Table~\ref{tab:burst_detec_prop}). All of the bursts presented in \citet{gajjar_2018_apj} are re-detected and from the \citet{zhang_2018_apj} sample we re-detected $30/46$ of the candidate bursts, missing only low signal-to-noise (S/N) pulses (where the astrophysical nature of these events is in any case unclear). We find $19$ bursts that were missed in both previous searches of this dataset. Of the newly found bursts, $8$ are very short in duration, with the entire burst envelope lasting no more than about $15$\,$\upmu$s (Figures~\ref{fig:family_plot} and \ref{fig:dur_hist}). We refer to these bursts as `ultra-fast radio bursts', or ultra-FRBs, because of their microsecond durations.

Bursts B23, B26 and B32 extend up to about $8.4$\,GHz, making them the highest-frequency FRBs detected to date. The emission properties above this frequency are unknown because above $8.4$\,GHz there is a sharp drop in the sensitivity of the receiver. 

The distribution of burst durations (as they were found in the search, Figure~\ref{fig:dur_hist}) may appear by eye to be bimodal, but accounting for Poissonian uncertainties we find no clear statistical evidence for this. In fact, the burst durations are marginally consistent with being logarithmically uniform from microseconds to milliseconds. If we split the burst sample in two at a duration of $40$\,$\upmu$s, we find $8^{+12.7}_{-5.9}$ and $41^{+23.1}_{-16.6}$ bursts below and above this threshold, respectively. Here the uncertainties represent the $3\sigma$ Poissonian error.

To study the bursts with more precision, we created high-time-resolution, full-polarisation data products for a selection of bursts; these have been coherently dedispersed to a DM of $560.105$\,pc\,cm$^{-3}$ (Methods). These data products have a time-resolution of $341.\bar{3}$\,ns and a frequency resolution of $2.9296875$\,MHz ($\delta \nu \delta t = 1$, limited by the uncertainty principle). It would be difficult to accurately achieve a higher time resolution by inverting the signal processing step that formed the frequency channels. 

\subsection*{High-time-resolution}\label{sec:hightimeres}

\begin{figure*}
    \vspace*{-3.1cm}
    \centerline{\includegraphics[height=1.0\textheight]{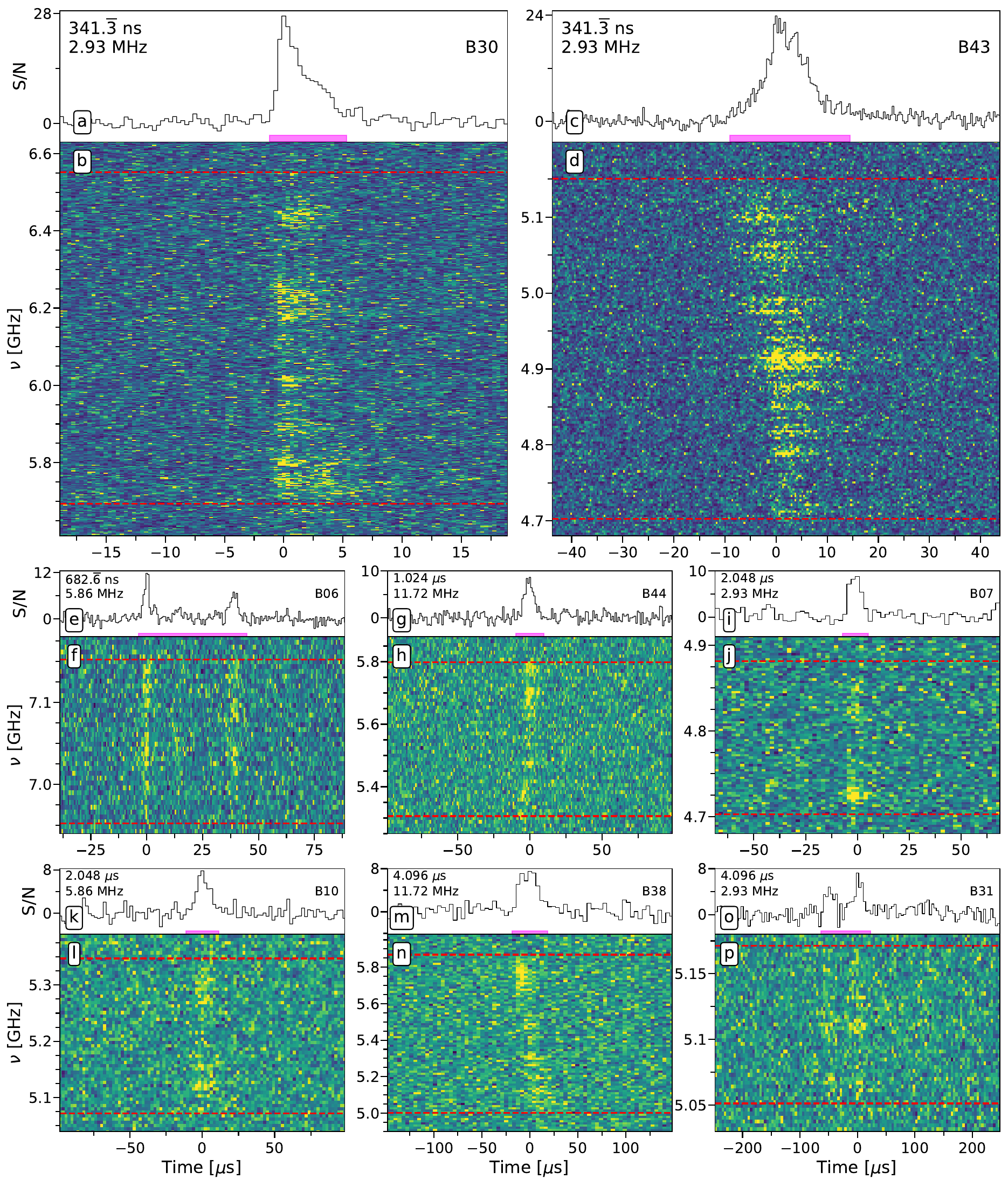}}
    \caption{\textbf{Total intensity dynamic spectra and profiles of the 8 ultra-FRBs.} We plot the dynamic spectrum (bottom panels) of the total-intensity data (Stokes~I) and their corresponding frequency-averaged time series (upper panels) for the 8 bursts with a detection duration $<30$\,$\upmu$s, sorted by their peak S/N. The time and frequency resolutions of the dynamic spectra are indicated in the top-left corners and the burst identifiers are shown in the top-right corners. The dashed red lines indicate the frequency range that was averaged over to create the burst profiles in the upper panels and the magenta bars indicate the time range over which the profile was integrated to determine the fluence of each burst (Table~\ref{tab:ufrbs_properties}). All the bursts have been coherently dedispersed to a DM of $560.105$\,pc\,cm$^{-3}$ (Methods). For visual purposes the limits of the colour map have been capped at the 99$^\mathrm{th}$ percentiles of the dynamic spectrum.}
    \label{fig:family_plot}
\end{figure*}

In Figure~\ref{fig:family_plot} we present the total intensity (Stokes~I) dynamic spectra and burst profiles for the $8$ shortest-duration bursts in our sample. We coherently dedisperse all of the bursts to the best-fit DM for burst B30 (Methods and Extended Data Figures~\ref{fig:peaksn} and \ref{fig:b30b43_comp}), $\mathrm{DM}=560.105$\,pc\,cm$^{-3}$, because of its high S/N and extremely narrow burst width. Nonetheless, we note that there is unavoidable ambiguity in the exact DM of each burst because their time-frequency morphology is not known {\it a priori}.

Bursts B30 (Figure~\ref{fig:family_plot}ab) and B43 (Figure~\ref{fig:family_plot}cd) are the highest-S/N bursts in our ultra-FRB sample. Like other bursts detected in these observations, B30 shows bright patches in frequency (consistent with the expected Galactic scintillation of $8$\,MHz at $4.7$\,GHz and $87$\,MHz at $8.0$\,GHz\cite{gajjar_2018_apj}), and is extended over more than $800$\,MHz of bandwidth. The leading edge of the burst is extremely steep, going from the noise floor to the peak of the burst in about $1$\,$\upmu$s. The entire burst lasts a mere $5$\,$\upmu$s, making it the shortest isolated FRB to date.

Burst B43 is comparatively long-duration: about $15$\,$\upmu$s. The shape of the burst in time-frequency space remains curved after dedispersing: the remaining curvature does not appear to follow the $\nu^{-2}$ relation for dispersion and therefore cannot simply be corrected for using a different DM. Compared to B30, it also shows clear bright patches in frequency, but a less steep leading edge. Rather, the profile of the burst (Figure~\ref{fig:family_plot}c) is more Gaussian shaped. Burst B43 is also found in the \citet{zhang_2018_apj} search due to the high S/N of the burst, but since their data were not coherently dedispersed and had a sampling time of $350$\,$\upmu$s (much larger than the total burst width), the microsecond duration of this burst was not visible.

Burst B06 (Figures~\ref{fig:family_plot}ef and \ref{fig:ilv_family}ef) shows multiple components, and is the only ultra-FRB that is detected at a central radio frequency $\geq7$\,GHz. The burst is composed of three components with separations of roughly $15$\,$\upmu$s and $25$\,$\upmu$s. The first component is the brightest and has a duration of only about $4$\,$\upmu$s.

The remaining ultra-FRBs are weaker and need to be downsampled in time and/or frequency to be clearly seen in the dynamic spectra. The spectral extent of the bursts varies between about $120$\,MHz (burst B31) and $800$\,MHz (bursts B30 and B38). The ultra-FRBs do not obviously favour any particular frequency-range nor do they occupy a particular time-range compared to the other bursts in this dataset (Extended Data Figures~\ref{fig:dur_vs_time} and \ref{fig:spec_vs_time}).

We calculate the peak flux density, fluence and isotropic-equivalent spectral luminosity of the 8 ultra-FRBs (Methods and Table~\ref{tab:ufrbs_properties}). We also show these in the context of the transient phase space diagram (Extended Data Figure~\ref{fig:tps}). The bursts all have a relatively high spectral luminosity and low transient duration compared to other bursts from \rone\ and other localised repeating FRBs. Notably, burst B30 has an inferred brightness temperature that exceeds $10^{40}$\,K. Such high brightness temperatures have so far only been observed in `nanoshots' from the Crab pulsar\cite{hankins_2003_natur,hankins_2007_apj,jessner_2010_aa} and in sub-bursts from FRB~20200120E\cite{nimmo_2022_natas}.

With the refined DM of $560.105$\,pc\,cm$^{-3}$, measured using B30 (Methods), we carefully examined the three highest-S/N bursts discovered in \citet{gajjar_2018_apj} on microsecond timescales (Figure~\ref{fig:ilv_family}). No micro-structure was identified in any of these bursts. A more detailed study of the larger sample at high time-resolution will be presented in a future paper. 

\subsection*{Polarimetric properties}\label{sec:pol}

After calibrating the full polarisation data (Methods), we measure the RM of our burst sample using a multi-burst joint Stokes~QU-fit\cite{michilli_2018_natur} (Figure~\ref{fig:multi_rm}). Similar to an analysis of bursts from \rone\ from the 305-m Arecibo telescope by \citet{michilli_2018_natur}, we find that all of the bursts, including the ultra-FRBs, can be fit with a single polarisation position angle (PPA), when averaged over the burst duration and an RM of $93586 \pm 4$\,rad\,m$^{-2}$ (in the observer frame), which is consistent with previous RM measurements of these bursts \cite{michilli_2018_natur, gajjar_2018_apj, faber_2021_rnaas}. We de-Faraday the Stokes parameters Q and U using the measured RM and determine the unbiased linear polarisation fraction, $\mathrm{L}_{\mathrm{unbias}}$ (Methods). 

\begin{figure*}
    \vspace*{-2cm}
    \centerline{\includegraphics[width=1.2\textwidth]{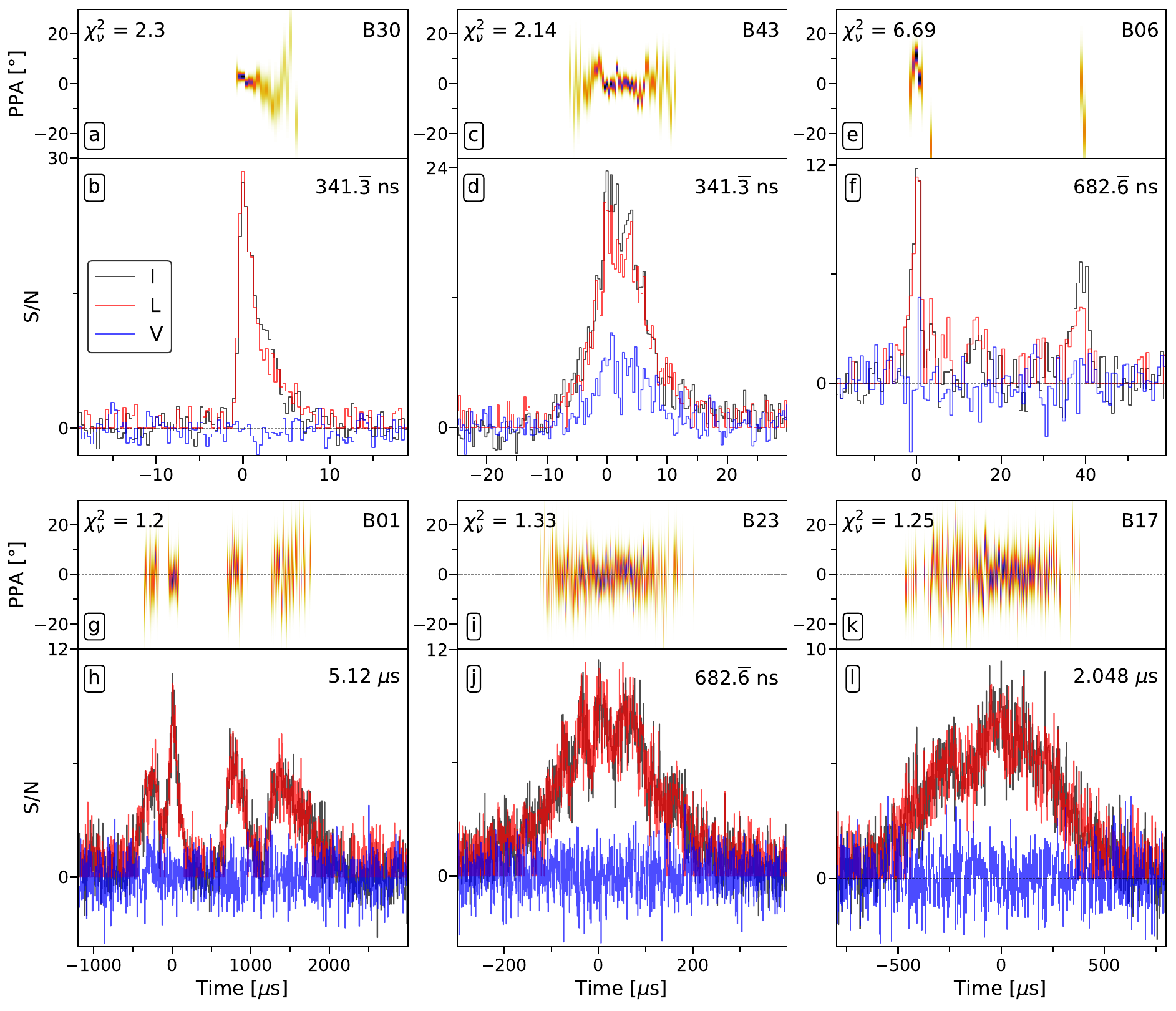}}
    \caption{\textbf{Full-polarisation, frequency-averaged profiles and polarisation position angles (PPAs) for a selection of bursts.} The top row shows the three brightest ultra-FRBs and the bottom row shows the three highest S/N millisecond-duration bursts (Supplementary Table~\ref{tab:burst_detec_prop}). The profiles of the bursts (bottom panels) show the total intensity (Stokes~I) in black, circular polarisation (Stokes~V) in blue and the Faraday-rotation-corrected unbiased linear polarisation in red (Methods). The colour-scale of the PPA (top panels) represents the probability distribution per time sample, with the darker colours representing a higher S/N in the unbiased linear polarisation ($\mathrm{L}_{\mathrm{unbias}}$). PPAs are only shown for time samples where $\mathrm{L}_{\mathrm{unbias}} \geq 4$. Per burst, the PPAs were shifted such that their weighted mean was centred around zero and they were fit with a constant line. The reduced-$\chi^{2}$-values, $\chi^{2}_{\nu}$, for those fits are displayed in the top left corners. Every burst has been coherently dedispersed to a dispersion measure of $560.105$\,pc\,cm$^{-3}$ (Methods).}
    \label{fig:ilv_family}
\end{figure*}

We plot the profiles of Stokes~I, $\mathrm{L}_{\mathrm{unbias}}$ and the circular polarisation (Stokes~V) in the bottom panels of Figure~\ref{fig:ilv_family}. We find that almost all of the ultra-FRBs are consistent with being $100$\,\% linearly polarised and show no circular polarisation --- similar to the millisecond-duration bursts\cite{michilli_2018_natur,gajjar_2018_apj,faber_2021_rnaas}. The exception to this is burst B43, which shows $31\pm10$\,\% circular polarisation and $87\pm10$\,\% linear polarisation: still consistent with being $100$\,\% polarised overall (Methods). 

We compute the time-resolved PPA across the burst profile. Due to the limitations of our calibration method (Methods), the absolute value of the PPA is not meaningful. We therefore, per burst, subtract the weighted mean, using $\mathrm{L}_\mathrm{unbias}$ as weights, so the PPA values are centred around zero. A constant line is then fit to the PPAs and we measure the reduced-$\chi^{2}$, $\chi^{2}_{\nu}$. The PPA probability distribution\cite{everett_2001_apj} for every time bin is shown in the upper panels of Figure~\ref{fig:ilv_family}. We find that at high time-resolutions $\chi^{2}_{\nu} > 1$, indicating that the PPAs are significantly scattered. This is also observed in a comparable analysis of the three selected bursts from the \citet{gajjar_2018_apj} sample (Figure~\ref{fig:ilv_family}).

Bursts B30 (the shortest-duration burst in our sample, Figure~\ref{fig:ilv_family}ab) and B01 (the longest-duration burst in our sample, Figure~\ref{fig:ilv_family}gh) are intentionally placed on top of each other to illustrate that, even though their durations differ by a factor of roughly $600$, their polarimetric properties are very similar. 

\section*{Discussion}\label{discussion}

Galactic neutron stars, including rotation-powered radio pulsars and radio-emitting magnetars, are known to produce at least three categories of coherent radio pulses\cite{lorimer_2012_hpa}: i. canonical pulsar emission from the magnetic polar cap; ii. microsecond-duration (or shorter) giant pulses from both energetic millisecond and young pulsars (like the Crab pulsar); and iii. magnetar radio bursts. The durations, spectra, rotational phase, etc., of these various emission types strongly suggest that they originate in different parts of the neutron star magnetosphere and likely via different physical processes. Here we have shown that FRB sources can also produce isolated microsecond-duration pulses, whose luminosities are many orders of magnitude brighter than even the giant pulses from the Crab pulsar (Extended Data Figure~\ref{fig:tps}). The range of timescales we observe from \rone\ is consistent with what is seen from neutron stars, and this may indicate that FRBs can also produce multiple burst types. The similarities between FRB microstructure and Crab giant pulses are discussed in more detail in, for example, \citet{nimmo_2022_natas}.

Microsecond (and shorter) timescale variations have previously been observed in bursts from FRB~20180916B \cite{nimmo_2021_natas} and FRB~20200120E \cite{nimmo_2022_natas}. In these cases, the short timescale variations occur within a broader burst envelope. In contrast, here we present the discovery of isolated microsecond-duration bursts (without a broader envelope) from \rone. These ultra-FRBs were missed in previous searches of the same data, where the time resolution was more than 100 times lower \cite{gajjar_2018_apj,zhang_2018_apj}. As suggested in \cite[Nimmo~et~al.][]{nimmo_2021_natas,nimmo_2022_natas}, we have shown here that there indeed exists a population of ultra-FRBs that current FRB search strategies are missing. The computational expense of searching for ultra-FRBs, because of the required high time resolution and coherent dedispersion, has resulted in them being undetected in all other FRB searches to date. Furthermore, scattering is a significant limitation to resolving such timescales when observing at low radio frequencies (less than about $1$\,GHz).

While FRB~20180916B\cite{marcote_2020_natur, tendulkar_2021_apjl}, FRB~20200120E\cite{kirsten_2022_natur} and \rone\cite{chatterjee_2017_natur,tendulkar_2017_apjl,marcote_2017_apjl,bassa_2017_apjl} live in drastically different environments, their burst properties show striking similarities. For example, similarities in the burst morphologies and polarimetric properties, including downward-drifting structure in time and frequency, spectrally narrow emission\cite{hessels_2019_apjl, chime_2019_apjl,nimmo_2023_mnras}, and now microsecond-duration fluctuations as well\cite{nimmo_2021_natas,nimmo_2022_natas}. In addition to the short timescales, the 3 orders-of-magnitude range of timescales observed in this work resembles the range of timescales observed for FRB~20180916B\cite{nimmo_2021_natas} and FRB~20200120E\cite{nimmo_2022_natas}. However, it is possible that microsecond-duration bursts are equally, or even more common in non-repeating FRBs\cite{cho_2020_apjl} due to them having narrower widths in general\cite{pleunis_2021_apj}. That micro-shots have so far been exclusively detected in repeaters is likely because of the high-time-resolution follow-up observations conducted for these sources. Alternatively, it could also be because more individual pulses have been seen and studied in the case of repeating FRBs. There are close to $500$ one-off FRBs in the first CHIME/FRB catalogue\cite{chime_2021_apjs}, but they lack the required time resolution to search for these short-duration sub-bursts. Additional studies using CHIME/FRB baseband detections of these FRBs will help elucidate the prevalence of micro-structures in one-off FRBs.

The polarimetric properties of repeating FRBs are often described by high degrees of linear polarisation, little-to-no circular polarisation, flat PPAs during the burst duration and only small ($<30^{\circ}$) variations in PPA between bursts detected at the same observing epoch\cite{michilli_2018_natur,nimmo_2021_natas,nimmo_2022_natas}. FRB~20180916B and \rone\ have also been seen to depolarise with decreasing frequency\citep{pleunis_2021_apjl,plavin_2022_mnras,feng_2022_sci,beniamini_2022_mnras}. In this work, we additionally show that, at very high time-resolution, the PPA during the burst duration varies significantly (Figure~\ref{fig:ilv_family}), resembling the high-time-resolution PPA variations seen for FRB~20180916B\cite{nimmo_2021_natas} and FRB~20200120E\cite{nimmo_2022_natas}. 

Recently, \citet{feng_2022_scibu} showed that \rone\ bursts at $1.25$\,GHz are occasionally circularly polarised (in $<1$\,\% of bursts) while being $0$\,\% linearly polarised (as is the case for all \rone\ bursts at these frequencies). We observe one ultra-FRB in our sample to be significantly circularly polarised, burst B43: this corresponds to a rate of circular polarisation for $4$--$8$\,GHz of a few percent which appears to be roughly consistent with \citet{feng_2022_scibu}, though a larger sample of high-radio-frequency bursts should be considered in a future study to compare the rates. \citet{feng_2022_scibu} suggest that Faraday conversion or the radiation mechanism itself could be responsible for the circular polarisation, while multi-path propagation in either the Milky Way or FRB local environment is disfavoured. Our detection of circular polarisation at much higher radio frequencies supports the idea that the origin is intrinsic to the emission mechanism, as opposed to being a propagation effect; we speculate that the rare occurrence of circular polarisation could be a result of how the emission is beamed towards Earth. Furthermore, circular polarisation has been detected from the repeating FRB, FRB~20220912A, which has been shown to live in a relatively clean local environment, with no changes in DM or RM on month timescales \cite{zhang_2023_arxiv,feng_2023_arxiv}. This also suggests that the circular polarisation is intrinsic to the emission mechanism.

The distribution of burst durations (Figure~\ref{fig:dur_hist}) is marginally consistent with being log-uniform between microseconds to milliseconds. The selection bias as a function of burst duration is not well known, and could swing either in favour of more ultra-FRBs or more typical-duration FRBs. In our search, the timescales below $30$\,$\upmu$s have only been searched with one trial boxcar, while timescales $\geq30$\,$\upmu$s have effectively been searched at least twice due to different combinations of the sampling time and boxcar matched-filtering widths (Figure~\ref{fig:dur_hist}). If all bursts from \rone\ draw their fluences, $\mathcal{F}$, from the same underlying distribution, ultra-FRBs should actually be easier to find since the detection metric is $\mathcal{F}/\sqrt{W}$, with $W$ the temporal duration of the burst. It is unclear, however, whether the micro-shots follow the same fluence distribution as the wider bursts.

The burst spectra and polarimetric properties (polarisation fractions, RM, and PPA) are consistent between the ultra-FRBs and the wider-bursts, suggesting that they share a common emission mechanism. It is unclear, however, why this mechanism sometimes shuts off after only microseconds whereas at other times it continues for several milliseconds. For \rone, the distribution of timescales is roughly uniform from microseconds to milliseconds. Other sources may show different temporal distributions and average burst duration; in any case, FRB searches are currently highly biased against finding ultra-FRBs. FRB~20200120E's bursts are typically 100\,$\upmu$s in duration\cite{nimmo_2023_mnras}, much shorter than other repeating FRBs\cite{pleunis_2021_apj}, further supporting the argument for an undetected FRB population with shorter timescales. 

In \citet{nimmo_2023_mnras} approximately $50$ bursts from FRB~20200120E are analysed but only one burst has clear microstructure superimposed on a broader envelope. For that one burst the ratio between the amplitude of the micro-shot and the broader envelope is only about $3$ --- meaning that it would require fine-tuning of the telescope sensitivity to detect only the micro-shot and not the broader emission as well. For both FRB~20200120E and FRB~20180916B the broader bursts were found in searches with a time resolution of $64$\,$\upmu$s\cite{nimmo_2022_natas} and $82$\,$\upmu$s\cite{marcote_2020_natur}, respectively. The micro-shots in those data were only found once the baseband data were both coherently dedispersed and higher time resolution data products were generated. At $64$\,$\upmu$s time resolution the ultra-FRBs from \rone\ would not have been found. This supports the idea that the ultra-FRBs from \rone\ are truly isolated events.

Fluctuations on timescales of tens of nanoseconds have been observed within the envelopes of bursts originating from FRB~20200120E\cite{nimmo_2022_natas}. With the current data, we are unable to probe similar timescales in the \rone\ ultra-FRBs and wider-bursts. The data used in this study has a limiting time-resolution of $341.\bar{3}$\,ns and for the majority of the ultra-FRBs discovered in this work the S/N is relatively low. The large scatter observed in the PPA at high time-resolution, however, is suggestive of the bursts being made up of narrower shots of emission. The 3 wider bursts from the \citet{gajjar_2018_apj} sample that we also study here show no evidence of micro-structure within the burst envelopes (Figure~\ref{fig:ilv_family}). This supports the hypothesis that the ultra-FRBs in our sample are truly isolated events as opposed to being fluctuations within a wider, dimmer burst component.

The isolated ultra-FRBs thus also provide a better constraint on the size of the emission region because they are unlikely to be generated by propagation effects that modulate the burst brightness post-emission\cite{beniamini_2020_mnras_498}. The extremely short duration of B30 constrains the emission region to be smaller than a few kilometres (ignoring any potential relativistic effects). This favours models in which the bursts are generated near the central engine in a magnetosphere - as opposed to models in which the bursts are generated much further out in a relativistic shock.

The existence of ultra-FRBs could influence what one infers about energy, wait time and burst rate distributions. We therefore encourage that future experiments should try to search the widest-possible range of timescales. This should especially be done for repeating FRBs where the known DM and just a single pointing direction make this computationally easier compared to an untargeted wide-field search.

The limitations of extremely high-time-resolution FRB searches include: knowing the DM {\it a priori} to be able to coherently dedisperse the data, or alternatively performing a search using both coherent and incoherent dedispersion steps; scattering from the Milky Way interstellar medium, which causes temporal broadening that increases with decreasing frequency; instrumental bandwidth limiting the sampling rate of the observations; and, simply, the overall computational cost. 
The discovery of ultra-FRBs from \rone\ in this work highlights that such transients exist, that they are being missed in current FRB searches, and that there is reason to invest time in overcoming the challenges associated with searches for ultra-FRBs in the future, particularly at giga-Hertz frequencies where scattering from the Galactic interstellar medium is less than a microsecond for most directions on the sky. Opening a wider range of timescales for FRB discovery can both lead to the identification of new FRB source types as well as even more precise probes of the intervening magnetised plasma and gravitational potential\cite{petroff_2022_aarv}.

% Methods
\section*{Methods}\label{sec:methods}

\subsection*{The observation}\label{sec:obs}

Observations of \rone\ were conducted with the Robert C. Byrd Green Bank Telescope (GBT) using the $4$--$8$\,GHz (C-band) receiver on 2017 August 26 during a $6$-hour observing block. The first hour was used for calibration procedures: a noise diode scan was used for flux and polarisation calibration and a $5$-minute observation of the pulsar PSR~B0329+54 was performed to verify the calibration procedure. The last $5$\,hours of the observing block were split into $30$-minute sections and were spent on \rone\citep{gajjar_2018_apj}. Observations were conducted with the Breakthrough Listen (BL) digital backend \citep{macmahon_2018_pasp}, which recorded $8$-bit raw voltage (baseband) data from $3.9$\,GHz to $9.3$\,GHz --- fully covering the C-band receiver bandwidth. The analog down-conversion system provided four $1500$-MHz passbands configured with central frequencies of $8563.964844$, $7251.464844$, $5938.964844$, and $4626.464844$\,MHz, respectively. Each dual-polarisation (linear basis) passband was Nyquist-sampled using $8$-bit digitizers, polyphase channelized to $512$ `coarse' frequency channels, and distributed to a cluster of compute nodes that recorded these data to disk. In total there are $32$ compute nodes, $8$ for every passband. Every compute node recorded $187.5$\,MHz of bandwidth over $64$ `coarse' channels. The passbands have exactly $187.5$\,MHz of overlap with adjacent passbands (Supplementary Table~\ref{tab:nodes}). Due to the polyphase filter, the highest time resolution possible for these data is $\left(1500\text{\,MHz} / 512\right)^{-1} = \left( 2.9296875 \text{\,MHz} \right)^{-1} = 341.\overline{3}$\,nanoseconds, where the $\overline{3}$ represents a repeating digit. The nodes store the data in chunks to keep file sizes manageable. Every chunk consists of exactly $2^{26}$ time samples, corresponding to $2^{26} \times 341.\overline{3}$\,ns\,$= 22.91$\,seconds, and is $17.2$\,GB in size. Each $30$-minute scan thus consists of $78$ `full' chunks and one chunk containing the last $13.29$\,seconds. The chunks are stored in GUPPI (Green Bank Ultimate Pulsar Processing Instrument) raw format (i.e., \texttt{.raw} files). Detailed information of the observation, the BL project, the BL backend and the GUPPI raw format can be found in Refs. \cite{gajjar_2018_apj}, \cite{worden_2017_acaau}, \cite{macmahon_2018_pasp} and \ \cite{lebofsky_2019_pasp}, respectively. 

\subsection*{Data preparation}\label{sec:dataprep}
The baseband data (\texttt{.raw} files) of the first $30$\,minutes were downloaded from the BL open data archive. The data of $24$ different nodes (see Supplementary Table~\ref{tab:nodes} for a list of used and available nodes) was retrieved, covering the frequency range $3876.464844$\,MHz -- $8376.464844$\,MHz. In total, $79 \times 24 \times 17.2 \text{ GB} = 32.4$\,terabytes (TB) of \texttt{.raw} data was downloaded. For every chunk, the $24$ \texttt{.raw} files were spliced together in frequency using a modified version of \texttt{splicer\_raw.py}, which is a \texttt{Python} script that is part of the BL \texttt{extractor} package. The $79$ spliced \texttt{.raw} files, each $412$\,GB in size (the last one is $239$\,GB), were converted to a \texttt{Sigproc}\cite{lorimer_2011_ascl} filterbank file using \texttt{digifil}. This is the \texttt{digifil} that is part of \texttt{bl-dspsr}. The filterbank files have been coherently dedispersed within the channels, but not between the channels, using a dispersion measure (DM) of $560.5$\,pc\,cm$^{-3}$ (from \citet{hessels_2019_apjl}). The resulting filterbank files contain $8$-bit intensity data and have a time-resolution of $341.\overline{3}$\,ns. The $4500$-MHz bandwidth of the files is made out of $1536$\,channels with a frequency resolution of $2.9296875$\,MHz. 

\subsection*{The search}\label{sec:search}

Benchmarking showed that searching the data at a time resolution of $341.\overline{3}$\,ns was too computationally expensive. Instead we opted to further downsample the data in time, using \texttt{digifil} to create new \texttt{Sigproc} filterbanks at time resolutions of $2.048$, $32.768$ and $524.288$\,$\upmu$s, respectively.

Previous searches\cite{gajjar_2018_apj} of these data showed that the bursts are narrowband. Therefore we create multiple `subbands' out of the $4.5$-GHz bandwidth filterbank file. We use subbands of $1500$, $750$, $375$ and $187.5$\,MHz, where each subband has a $50$\,\% overlap with adjacent subbands, e.g., the $4500$-MHz file is split into five $1500$-MHz files. A total of 
\begin{equation}
\begin{split}
        \ & 1 + \sum_{i} 2 \times 4500 / i - 1 = 87 \\
    \ & \mathrm{where } \ \ i \in \left[ 1500, 750, 375, 187.5 \right]
\end{split}
\end{equation}
unique subbands were searched. Combined with the three possible time-resolutions of the data, there were $87 \times 3 = 261$ passes through the data.

Every filterbank file was dedispersed and summed over all frequency channels using a range of dispersion measures (Supplementary Table~\ref{tab:prepsubband}) with \texttt{PRESTO}'s \texttt{prepsubband}. The steps in DM are equal, or smaller, than the step-size suggested by \texttt{PRESTO}'s \texttt{DDplan.py}. The resulting time series were searched with \texttt{PRESTO}'s \texttt{single\_pulse\_search.py} using a S/N threshold of $7$. For the $2$\,$\upmu$s and $33$\,$\upmu$s data we made use of all the available boxcar widths, which are logarithmically scaled between $1$ and $300$, while for the $524$\,$\upmu$s data the highest boxcar width that was used is $45$. This made the search sensitive to timescales between $2$\,$\upmu$s and $24$\,ms -- which is the maximum duration of a large sample of \rone\ bursts at about $1.4$\,GHz as presented in \citet{hewitt_2022_mnras}. At no stage do we apply any radio frequency interference (RFI) excision, to avoid masking any potential bright or short-duration bursts.

All the candidates were classified using the machine learning algorithm \texttt{FETCH}\cite{agarwal_2020_mnras} (model \texttt{A}). We reject all the candidates that have a temporal width $>0.5$\,ms and probability of being an astrophysical signal $<50$\,\%. Since \texttt{FETCH} is trained on (simulated) FRBs with widths between $0.5$\,ms and $50$\,ms, we chose to manually inspect every candidate with a width $\leq0.5$\,ms, regardless of their \texttt{FETCH}-derived probability of being an astrophysical signal. 

We manually inspect the roughly $2300$ candidates that meet our selection criteria and find that about $2100$ are FRBs. Due to the complex time-frequency structure of some of the FRBs and the fact that we have $261$ passes over the data, all of the FRBs are found multiple times --- the most extreme case is burst B01, which is found over $400$ times (Supplementary Table~\ref{tab:burst_detec_prop}).

Bursts are sorted by their arrival times and are clustered using their proximity in time. A `new' cluster starts if the time difference between consecutive bursts exceeds $10$\,ms. Every cluster is manually checked for any potential bursts that are close enough in time that they were grouped together. We keep the clusters if at least one of the burst detections in that cluster has S/N $\geq 8$.

Given the discovery of the ultra-FRBs, we re-checked all the candidates with a duration of $2$\,$\upmu$s, regardless of their S/N or \texttt{FETCH} probability. After an additional manual inspection, none of these candidates are believed to be real astrophysical signals.

\subsection*{Dispersion measure determination}\label{sec:dmdetermination}

We use burst B30 to determine the best DM. This burst was chosen because it has a high S/N, extremely short duration and compared to the other microsecond-duration bursts it has a relatively large extent in frequency (Figure~\ref{fig:family_plot}b and Extended Data Figure~\ref{fig:b30b43_comp}bd). We coherently dedisperse the burst to a DM of $560.1$\,pc\,cm$^{-3}$ and incoherently dedisperse it over a range of trial DMs --- from $559.7$\,pc\,cm$^{-3}$ to $560.7$\,pc\,cm$^{-3}$, in steps of $0.0002$\,pc\,cm$^{-3}$. For every trial DM we sum the dynamic spectrum over the frequency extent $5650$--$6525$\,MHz. The profile of the burst is normalized such that the off-burst regions have zero mean and unit standard deviation and the peak S/N is recorded. We plot the results in Extended Data Figure~\ref{fig:peaksn}. Individual measurements are plotted as grey circles and for visual purposes a moving average is shown as a solid black line. The peak S/N value is not clearly maximized at one specific DM value. We attribute this to brightness fluctuations (varying in both time and frequency) of the burst in the dynamic spectrum (Figure~\ref{fig:family_plot}b and Extended Data Figure~\ref{fig:b30b43_comp}bd), as shown in e.g. \citet{hessels_2019_apjl}. As discussed there, this could be a radius-to-frequency mapping effect or, in principle, such effects could also arise if the emission at different frequencies is occurring at different physical distances from Earth. The light travel distance of $341.\bar{3}$\,ns (the highest possible time resolution) corresponds to about $100$\,m (in the absence of relativistic effects). At such high time resolution, subtle changes in the location of the emission region may become apparent. The frequency-dependent pulse shapes of pulsars are also known to complicate DM determination\cite{pennucci_apj_2014, pennucci_apj_2019}.

To determine the best DM, we fit a Lorentzian distribution to the data points close to the peak (shown as a solid green line) and take the best DM as the center of the fitted distribution, which is at $560.105$\,pc\,cm$^{-3}$. Due to the complex shape of the peak S/N profile we estimate the error on the DM to be $0.05$\,pc\,cm$^{-3}$. 

The same method is repeated for burst B43 (over the frequency extent $4700$--$5150$\,MHz). Extended Data Figure~\ref{fig:peaksn} shows individual measurements as grey pentagons and the dashed black line shows the corresponding moving average. All the individual data points are fitted with a Gaussian distribution, which peaks at a DM of $560.308$\,pc\,cm$^{-3}$. We estimate the error on this DM to be $0.25$\,pc\,cm$^{-3}$, given the relatively flat peak S/N profile of B43. Extended Data Figure~\ref{fig:b30b43_comp} illustrates the effects of using a different DM for bursts B30 and B43. A DM of $560.308$\,pc\,cm$^{-3}$ would decrease the duration of burst B43 by about $2$\,$\upmu$s (Extended Data Figure~\ref{fig:b30b43_comp}efgh). However, if a DM of $560.308$\,pc\,cm$^{-3}$ is applied to burst B30 it is clearly over-dedispersed (Extended Data Figure~\ref{fig:b30b43_comp}cd) and we therefore continue the analysis with a DM of $560.105$\,pc\,cm$^{-3}$.

We note that this DM is lower than the one reported in \citet{gajjar_2018_apj} ($565.0$\,pc\,cm$^{-3}$) and \citet{hessels_2019_apjl} ($563.86$\,pc\,cm$^{-3}$), both of which used a structure-maximizing DM on burst B01 (called `GB-BL' in \citet{hessels_2019_apjl} and `11A' in \citet{gajjar_2018_apj}). However, it should be mentioned that, using bursts between $1.2$\,GHz and $2.3$\,GHz, \citet{hessels_2019_apjl} find a DM of $560.5$\,pc\,cm$^{-3}$ for \rone\ bursts in general near the epoch of our GBT observations --- giving a better handle on the determined DM because of the lower radio frequencies.

\subsection*{Residual temporal smearing}

Even though the availability of voltage data allows for coherent dedispersion there could still be intra-channel smearing due to the usage of an incorrect value of the DM. Furthermore, intrinsically narrow pulses could also be broadened by scattering from the Milky Way's interstellar medium.

In \citet{gajjar_2018_apj} they used the Galactic electron density model \texttt{NE2001}\cite{cordes_2002_arxiv} and estimated a Galactic scattering timescale for \rone\ of $\tau_{s} = 20 \ \upmu \mathrm{s} \ \nu^{-\alpha}$. Here $\nu$ is the observing frequency in GHz and $\alpha$ is a scaling parameter which is between $4$ (thin screen model\cite{scheuer_1968_natur}) and $4.4$ (Kolmogorov spectrum\cite{lee_1975_apj}). We find that the scattering timescale is between $2$\,ns (best case, $\nu = 8.0$\,GHz and $\alpha = 4.4$) and $41$\,ns (worst case, $\nu = 4.7$\,GHz and $\alpha = 4.0$). Since the highest-possible time resolution available in our data is $341.\bar{3}$\,ns, temporal smearing due to Galactic scattering is expected to be undetectable in the temporal profiles of the bursts.

The dispersive sweep can be calculated as\cite{lorimer_2012_hpa}:
\begin{equation}
    \Delta t = \mathfrak{D} \times (\nu_{1}^{-2} - \nu_{2}^{-2}) \times \mathrm{DM}
\end{equation}
where $\mathfrak{D}$ is the dispersive constant, and $\nu_{1,2}$ are the upper- and lower bound of the frequency range. Thus, for coherent dispersion, a channel width of $2.9296875$\,MHz, observing frequencies of $4.7$\,GHz and $8.0$\,GHz and a dispersive constant of $1/(2.41 \times 10^{-4})$\,MHz$^{2}$\,pc$^{-1}$\,cm$^{3}$\,s (the same constant that is used in, e.g., \texttt{dspsr}) we find a smearing of $234$\,ns and $48$\,ns per unit DM within an individual channel. Since the error on the DM is about $0.05$\,pc\,cm$^{-3}$ (Extended Data Figure~\ref{fig:peaksn}) and the highest-possible time resolution is $341.\bar{3}$\,ns, temporal smearing due to dispersion within individual channels is also insignificant. 

\subsection*{Sensitivity to timescales}\label{sec:timescales}

The highest time resolution at which the data was searched was $\Delta t = 2.048$\,$\upmu$s with a S/N $\geq 8$ detection limit. Any bursts shorter than $\Delta t$ would thus have a decreased S/N since noise would have been added into the single time bin in which the burst would occur. The decrease in S/N scales with the square root of the fraction between the sampling time and the burst duration. Therefore it is possible that relatively weak and even shorter duration (less than $2$\,$\upmu$s) bursts are missed in the search. However, individual time samples in the profiles of bursts B30 and B43 reach a S/N greater than $20$ at a time resolution of $341.\bar{3}$\,ns (Figures~\ref{fig:family_plot}ac and \ref{fig:ilv_family}bd). A burst with a duration of $341.\bar{3}$\,ns and a S/N of $20$ would have its S/N decreased by a factor $\sqrt{6}$ and would appear as a S/N $20 / \sqrt{6} = 8.2$ candidate and would likely have been found in the search, though just barely.

\subsection*{Checks for saturation}\label{sec:saturation}
Bright FRBs can saturate the receiver system and/or recording backend (see \citet{Ikebe_2023_PASJ} and \citet{kirsten_2023_arxiv} for examples). We investigated if burst B43 is saturated because this burst is relatively bright between $4.9$\,GHz and $4.93$\,GHz (Figure~\ref{fig:family_plot}d). The baseband data of the BL backend are stored as complex $8$-bit signed integers\cite{macmahon_2018_pasp}, meaning that both the real and the imaginary part utilize one byte to store their respective signed integers (i.e., an integer between $-128$ and $+127$, inclusive). The baseband data of burst B43 was loaded into \texttt{Python} and every sample for all subbands and both polarisation channels were checked. We found that all the integer values are within $\pm105$, suggesting that there is no saturation in the recording system. It is not surprising that the signal is not saturated: the intensity of the signal is spread over both polarisation channels and over both the real and imaginary parts. Furthermore, the signal is also spread out in time in the baseband data due to dispersive smearing within a subband ($>100$\,$\upmu$s for B43) and only with coherent dedispersion is the microsecond duration of these bursts revealed. The $8$-bit depth of the baseband data used here is deeper than cases in which saturation did occur ($4$-bit baseband depth for the burst shown in \citet{Ikebe_2023_PASJ} and $2$-bit baseband depth for the bursts shown in \citet{kirsten_2023_arxiv}).

\subsection*{Polarimetric calibration}\label{sec:polcal}

\begin{figure*}
    \centerline{\includegraphics[width=1.2\textwidth]{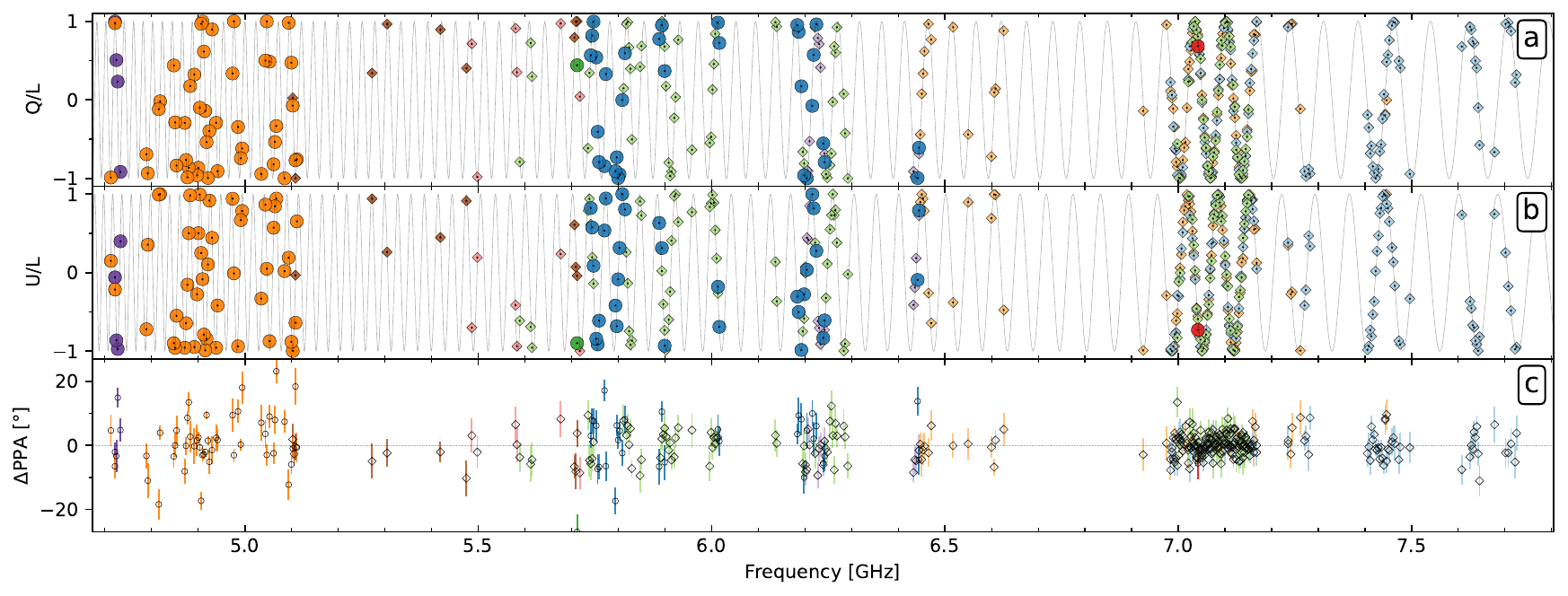}}
    \caption{\textbf{Multi-burst joint QU-fit.} The Stokes parameters Q (panel A) and U (panel B) oscillate due to Faraday rotation. The Stokes parameters have been normalised by the total measured linear polarisation ($\mathrm{L_{\mathrm{meas}}}=\sqrt{\mathrm{Q}^{2}+\mathrm{U}^{2}}$). Different colours represent different bursts. Diamonds represent bursts that were previously found by \citet{gajjar_2018_apj} and circles represent newly discovered microsecond-duration bursts. Panel C shows the difference between the measured and modeled PPAs and the $1\sigma$ uncertainty on each PPA measurement. Only data points with a S/N $\geq5$ are shown.}
    \label{fig:multi_rm}
\end{figure*}

To study the bursts in greater detail, we use \texttt{digifil} to coherently dedisperse (using a DM of $560.105$\,pc\,cm$^{-3}$) the baseband data for all the bursts that are shown in Figures~\ref{fig:family_plot}, \ref{fig:ilv_family} and \ref{fig:multi_rm}. The resulting files contain the coherence products and have a time resolution of $341.\bar{3}$\,ns, a frequency resolution of $2.9296875$\,MHz and a total bandwidth that varies per burst. The files are stored in filterbank\cite{lorimer_2011_ascl} format using $32$-bit floating-point numbers to avoid potential saturation effects.

In a linear basis with complex sampling, the Stokes~I, Q, U and V parameters can be constructed from the auto- and cross-correlations (i.e., the coherence products) using\cite{vanstraten_2010_pasa}
\begin{align}
    \mathrm{I} &= \left< \mathrm{AA}^{*} + \mathrm{BB}^{*} \right> \label{eq:si} \\
    \mathrm{Q} &= \left< \mathrm{AA}^{*} - \mathrm{BB}^{*} \right> \label{eq:sq} \\
    \mathrm{U} &= \left< 2 \ \mathrm{Re} \left( \mathrm{AB}^{*} \right) \right> \label{eq:su} \\
    \mathrm{V} &= \left< 2 \ \mathrm{Im} \left( \mathrm{AB}^{*} \right) \right> \label{eq:sv}
\end{align}
where $\mathrm{AA}^{*}$ and $\mathrm{BB}^{*}$ are the auto-correlations of the polarisation channels and $\mathrm{Re} \left( \mathrm{AB}^{*} \right)$ and $\mathrm{Im} \left( \mathrm{AB}^{*} \right)$ are the real and imaginary parts of the cross-correlations. 
An instrumental delay between the polarisation channels will affect the cross-correlation $\mathrm{AB}^{*}$, which in turn will affect the Stokes parameters U and V. We make use of the baseband data of a noise diode scan, taken $2$\,minutes before the start of the observation, to determine the delay between the polarisation channels. The data were folded on the switching period ($0.04$\,s) of the noise diode using \texttt{dspsr}\cite{vanstraten_2011_pasa}. An `archive' format file was made for each of the four passband containing the coherence products. We determine the phase angle (PA) between the cross-products using:
\begin{equation}
    \mathrm{PA} \left( \nu \right) = \frac{1}{2} \arctan{ \left[ \mathrm{Re} \left( \mathrm{AB}^{*} \right) / \ \mathrm{Im} \left( \mathrm{AB}^{*} \right) \right] }
    \label{eq:phase}
\end{equation}

A slope $\mathcal{S}$ is fit to the $\mathrm{PA}$, taking the wrapping at $\pm \pi/2$\,rad into account. The slope $\mathcal{S}$, which has units of $\mathrm{rad}/\mathrm{Hz}$, is converted to a delay $\mathcal{D}$ using $\mathcal{D} = \mathcal{S} / \left( \pi \  \mathrm{rad} \right)$.

The delay between the polarisation channels is, as expected, different per passband but is on the order of $2.5$\,nanoseconds --- similar to delays found in other radio telescopes\cite{mckinven_2021_apj, kirsten_2021_natas, nimmo_2021_natas}.

We correct the cross-correlations for the instrumental delay using:
\begin{align}
    \mathrm{Re} \left( \mathrm{AB}^{*} \right)_{\mathrm{corrected}} = \mathrm{Re} \left( \mathcal{Y} \right) \\
    \mathrm{Im} \left( \mathrm{AB}^{*} \right)_{\mathrm{corrected}} = \mathrm{Im} \left( \mathcal{Y} \right) \\
    \mathrm{where} \ \mathcal{Y} = \left[ \mathrm{Re} \left( \mathrm{AB}^{*} \right) + i \ \mathrm{Im} \left( \mathrm{AB}^{*} \right) \right] \times e^{-2i \pi \nu \mathcal{D}}
\end{align}

The Stokes parameters are constructed from the auto-correlations and the delay-corrected cross correlations using Equations\,\ref{eq:si}--\ref{eq:sv}. Every channel for every Stokes parameter is normalized by subtracting the mean and dividing by the standard deviation of an off-burst region of that channel. 

We determine the rotation measure (RM) using a multi-burst joint QU-fit (Figure~\ref{fig:multi_rm}) using the following equations:
\begin{align}
    \mathrm{Q} / \mathrm{L} = \cos & \left( 2 \left[ c^{2} \mathrm{RM} / \nu^{2} + \phi \right] \right) \label{eq:rmq} \\
    \mathrm{U} / \mathrm{L} = \sin & \left( 2 \left[ c^{2} \mathrm{RM} / \nu^{2} + \phi \right] \right) \label{eq:rmu}
\end{align}
where $c$ is the speed of light, $\mathrm{L}$ the quadrature sum of $\mathrm{Q}$ and $\mathrm{U}$, i.e. $\mathrm{L} = \sqrt{\mathrm{Q}^2 + \mathrm{U}^2}$ and $\phi = \phi_{\inf} + \phi_{\mathrm{inst}}$ where $\phi_{\inf}$ is the absolute angle of the polarisation of the sky referenced to infinite frequency and $\phi_{\mathrm{inst}}$ is the phase difference between the polarisation hands. In the fit, the parallactic angle is assumed to be the same for all the bursts. In reality the parallactic angle changes by $6^{\circ}$ between the first and the last burst in our dataset. We deviate from the more generalized form of these equations (see, e.g., Equations\,4 and 5 of \citet{nimmo_2021_natas}) that incorporate a term for the instrumental delay since we have removed the delay {\it a priori}. We find an observed RM, $\text{RM}_{\text{obs}}$, of $93586 \pm 4$\,rad\,m$^{-2}$, where the error is the $1\sigma$ statistical error. The RM in the source reference frame\cite{michilli_2018_natur}, $\text{RM}_{\text{src}}$, is $(1+z)^{2} \ \text{RM}_{\text{obs}} = 1.42 \ \text{RM}_{\text{obs}} = 1.3 \times 10^{5}$\,rad\,m$^{-2}$, where $z$ is the redshift of the host galaxy of \rone\cite{tendulkar_2017_apjl}. This RM is consistent with previously reported RM values of bursts from the same dataset\cite{michilli_2018_mnras,gajjar_2018_apj,faber_2021_rnaas}.

The dynamic spectrum of the Stokes parameters Q and U are corrected for Faraday rotation with the aforementioned RM, using: 
\begin{align}
    \mathrm{Q}_{\mathrm{corrected}} &= \mathrm{Re} \left( \mathcal{W} \right) \\
    \mathrm{U}_{\mathrm{corrected}} &= \mathrm{Im} \left( \mathcal{W} \right) \\
    \mathrm{where} \, \mathcal{W} &= \left[ \mathrm{Q} + i \, \mathrm{U} \right] \times e^{-2i \  c^{2} \nu^{-2} \ \mathrm{RM} }
\end{align}
We compute the time series of the Faraday-rotation-corrected Stokes~I, Q, U and V parameters by summing over the frequency extent of the bursts. Each of the four time series are then again normalized such that the off-burst regions have zero mean and unit standard deviation. We calculate the measured linear polarisation by taking the quadrature sum of Stokes~Q and U, $\mathrm{L}_{\mathrm{meas}} = \sqrt{\mathrm{Q}^2 + \mathrm{U}^2}$. Since $\mathrm{L}_{\mathrm{meas}}$ is derived from squared quantities it has a positive bias. To correct for this bias we follow the prescription as shown in \citet{everett_2001_apj}:

\begin{equation}\label{eq:Lunbias}
    \mathrm{L}_{\mathrm{unbias}}=\begin{cases}                               
    \sigma_{I}\sqrt{\left(\frac{\mathrm{L}_{{\mathrm{meas}}}}{\sigma_{I}}\right)^2-1}, & \text{if} \ \frac{\mathrm{L}_{{\mathrm{meas}}}}{\sigma_{I}}\ge 1.57\\                              
    0, & \text{otherwise}                                                               \end{cases}
\end{equation}

where $\sigma_{I}$ is the standard deviation in the off-burst Stokes~I.

The large channel widths of our very-high-time-resolution data and the extremely large RM of \rone\ causes the polarisation angle, $\theta$, to change significantly within one channel. The intra-channel Faraday rotation is given by\cite{michilli_2018_natur}
\begin{equation}
    \Delta \theta = \mathrm{RM}_{\mathrm{obs}} c^{2} \nu^{-3}_{c} \Delta \nu,
\end{equation}
where $c$ is the speed of light, $\nu_{c}$ is the observing frequency, and $\Delta \nu$ is the channel width. For an RM of $93586$\,rad\,m$^{-2}$ and a channel width of $2.9296875$\,MHz this results in rotations of $13.6^{\circ}$ and $2.8^{\circ}$ at $4.7$\,GHz and $8.0$\,GHz, respectively. The depolarisation fraction is given by\cite{michilli_2018_natur}
\begin{equation}
    f_{\mathrm{depol}} = 1 - \left[ \sin \left( 2 \Delta \theta \right) / \ 2 \Delta \theta \right],
\end{equation}
resulting in a depolarisation of $3.7$\,\% and $0.2$\,\% at these two frequencies, respectively.

We find that all but one of the bursts are consistent with being $100$\,\% linearly polarised (Figure~\ref{fig:ilv_family}). The sole exception is burst B43, which shows $87 \pm 10$\,\% linear polarisation and $31 \pm 10$\,\% circular polarisation (Figure~\ref{fig:ilv_family}d). We use a conservative $10$\,\% error which is a combination of the statistical error, systematic errors due to the instrumental calibration and the intra-channel depolarisation. Other bursts at similar frequencies as burst B43 show no signs of circular polarisation.

We determine the time-resolved PPAs across the burst profiles using:
\begin{equation}
    \mathrm{PPA} = \frac{1}{2} \arctan \left( \mathrm{U} / \mathrm{Q} \right).
    \label{eq:ppa}
\end{equation}
We compute the PPA for every sample where $\mathrm{L}_{\mathrm{unbias}} \geq 4$. For every burst we subtract the weighted mean of the PPAs, using $\mathrm{L}_{\mathrm{unbias}}$ as weights, and plot their probability distributions\cite{everett_2001_apj} in the upper panels of Figure~\ref{fig:ilv_family}. We fit the PPAs to a constant line by minimizing the weighted least-squares. The reduced $\chi^{2}$-value, $\chi^{2}_{\nu}$, is reported in Figure~\ref{fig:ilv_family} and often greatly exceeds $1$, indicating a significant scatter. 

\subsection*{Energetics}\label{sec:energies}

To determine the peak flux density, fluence and isotropic-equivalent spectral luminosity of the 8 ultra-FRBs we make use of the data products described in subsection `Polarimetric calibration'. First, the dynamic spectra of Stokes~I is averaged over a range of frequencies (as illustrated by the dashed horizontal red lines in Figure~\ref{fig:family_plot}). Next, the time series are normalized such that the off-burst regions have zero mean and unit standard deviation. To convert from S/N units to physical units we make use of the radiometer equation\cite{lorimer_2012_hpa} assuming a system temperature of $26$\,K and an antenna gain of $2$\,K\,Jy$^{-1}$, i.e. a system equivalent flux density (SEFD) of $13$\,Jy. These quantities are expected to have fractional uncertainties of at most $20$\,\%. To determine the fluence we integrate the profile over a specific time range that is indicated with a magenta bar in Figure~\ref{fig:family_plot}. The spectral luminosity is calculated assuming a luminosity distance of $972$\,Mpc\cite{tendulkar_2017_apjl} to \rone. The results are tabulated in Table~\ref{tab:ufrbs_properties} and plotted in Extended Data Figure~\ref{fig:tps}.

\section*{Data availability}

The data that support the plots within this paper and other findings of this study are available from \url{https://doi.org/10.5281/zenodo.8112803} or from the corresponding author upon reasonable request. The voltage data are available through the Breakthrough Initiatives Open Data Portal: \url{https://breakthroughinitiatives.org/opendatasearch}, and are explained in detail in: \url{http://seti.berkeley.edu:8000/frb-data/}.

\section*{Code availability}
The pulsar package \texttt{dspsr} is available at \url{https://dspsr.sourceforge.net/} and a modified version of \texttt{dspsr}, \texttt{bl-dspsr}, that is able to read voltage data from the Breakthrough Listen backend is available at \url{https://github.com/UCBerkeleySETI/bl-dspsr}. Code to splice and extract voltage data is available at \url{https://github.com/greghell/extractor}. \texttt{FETCH} can be found at \url{https://github.com/devanshkv/fetch}. The PRESTO suite of tools is available at \url{https://github.com/scottransom/presto}.

\section*{Acknowledgements}
{We would like to thank the Breakthrough Listen project for keeping the raw baseband data from these observations and making it publicly available. Breakthrough Listen is funded by the Breakthrough Initiatives (\url{https://breakthroughinitiatives.org/}). We thank the referees for their constructive comments that improved the manuscript. We thank J.~Weisberg for useful discussions about radio astronomy and polarimetry. A.D.~Seymour is thanked for tips regarding the GBT BL data. Research by the AstroFlash group at University of Amsterdam, ASTRON and JIVE is supported in part by an NWO Vici grant (PI Hessels; VI.C.192.045). K.N. is an MIT Kavli Fellow.}

\section*{Author contributions}
M.P.S. led the burst search, data analysis, and made the figures and tables. He wrote the majority of the manuscript. K.N. made significant contributions to the writing and provided guidance on the data analysis. J.W.T.H. supervised the work, guided the overall approach, and made significant contributions to the writing. All co-authors provided input on the scientific interpretation.

\section*{Competing interests}
The authors declare no competing interests.

\end{multicols}

\clearpage

\begin{table*}
\caption{\textbf{Burst properties of the 8 microsecond duration bursts.}}
\label{tab:ufrbs_properties}
\centering
\begin{tabular}{crrrccr}
\hline \hline
Burst & Peak S/N$^{\alpha}$ & Peak Flux Density$^{\alpha,\gamma}$ & Fluence$^{\gamma}$ & Spectral Luminosity$^{\gamma,\delta}$ & Frequency Extent$^{\zeta}$ & Time Extent$^{\psi}$ \\
 &  &  [Jy] &  [mJy ms] & [$10^{33}$\,erg\,s$^{-1}$\,Hz$^{-1}$] & [MHz] & [$\upmu$s] \\
\hline
B06 &  11.8 &   9.3 &   65 &  1.5 &  199.2 &  48.5 \\
B07 &   8.8 &   4.2 &   32 &  2.9 &  178.7 &  12.3 \\
B10 &   7.8 &   3.0 &   31 &  1.6 &  275.4 &  22.5 \\
B30 &  27.4 &  14.7 &   41 &  7.2 &  858.4 &   6.5 \\
B31 &   7.2 &   3.0 &   89 &  1.2 &  120.1 &  86.0 \\
B38 &   7.5 &   1.2 &   27 &  0.8 &  867.2 &  36.9 \\
B43 &  23.8 &  17.7 &  161 &  7.7 &  448.2 &  23.6 \\
B44 &   8.6 &   3.5 &   27 &  1.6 &  492.2 &  19.5 \\
\hline
\multicolumn{7}{l}{$^{\alpha}$ Determined for the time resolutions indicated in Figure~\ref{fig:family_plot}.} \\

\multicolumn{7}{l}{$^{\gamma}$ We estimate a conservative $20$\,\% error on these measurements, arising due to the uncertainty} \\
\multicolumn{7}{l}{\hspace{0.25cm} in the system equivalent flux density (SEFD) of the GBT.} \\
\multicolumn{7}{l}{$^{\delta}$ Isotropic-equivalent and using a luminosity distance of $972$\,Mpc\cite{tendulkar_2017_apjl}.} \\
\multicolumn{7}{l}{$^{\zeta}$ The frequency range over which the dynamic spectra was averaged,} \\
\multicolumn{7}{l}{\hspace{0.25cm} i.e., the difference between the red horizontal dashed lines in Figure~\ref{fig:family_plot}.} \\
\multicolumn{7}{l}{$^{\psi}$ The time range over which the profile was integrated to determine the fluence,} \\
\multicolumn{7}{l}{\hspace{0.25cm} i.e., the length of the magenta bar in Figure~\ref{fig:family_plot}.} \\

    \end{tabular}
\end{table*}

\clearpage

\clearpage

\clearpage

\clearpage

% Reset figure and table counters and rename them
\setcounter{figure}{0}
\captionsetup[figure]{name={Extended Data Figure}}

\setcounter{table}{0}
\captionsetup[table]{name={Supplementary Table}}

\clearpage

\begin{figure*}
    \centerline{\includegraphics[width=0.8\textwidth]{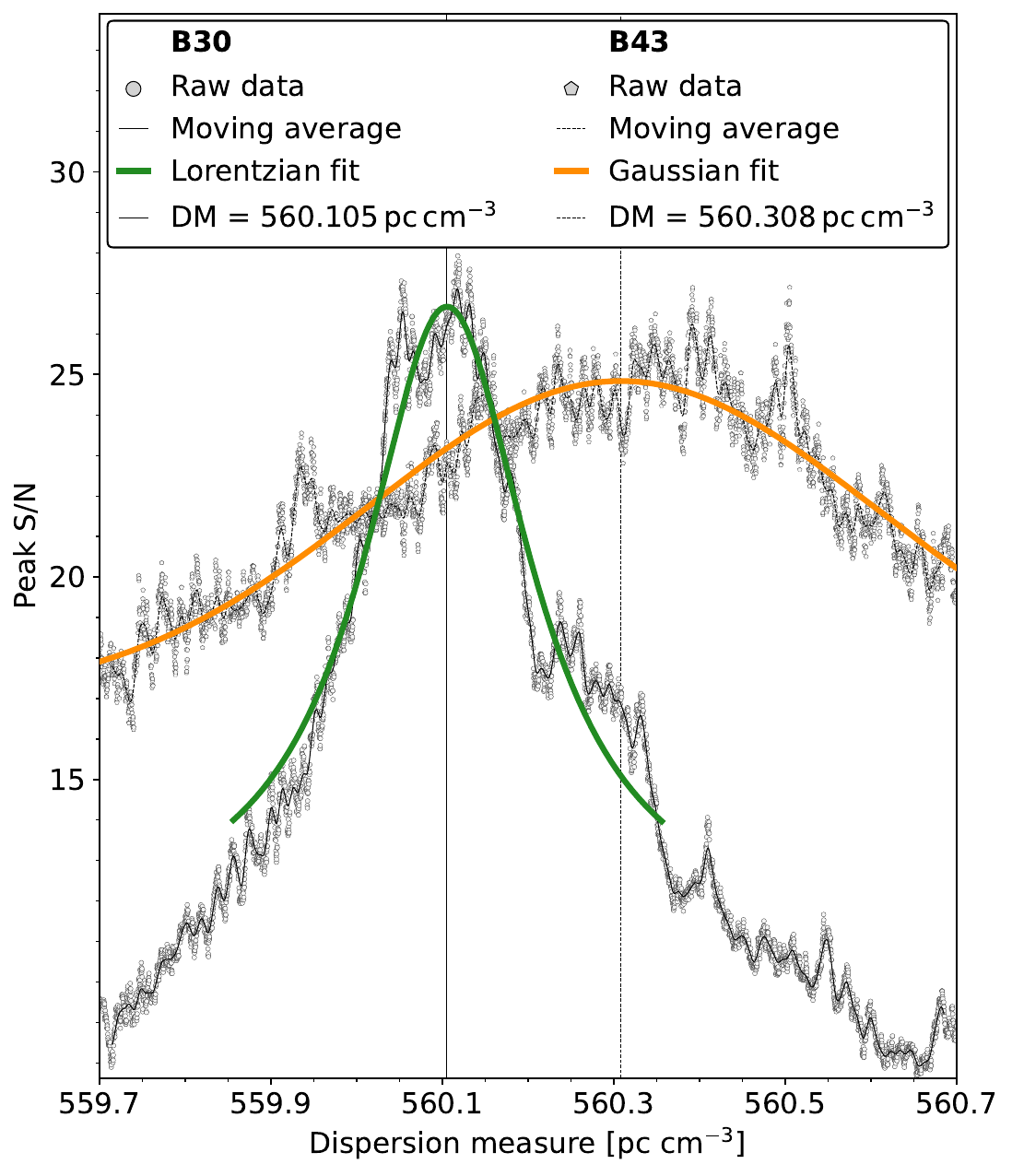}}
    \caption{\textbf{Peak S/N of the} $\mathbf{341.\bar{3}}$\textbf{-ns profiles of bursts B30 and B43, as a function of DM.} The bursts were first coherently dedispersed to a DM of $560.1$\,pc\,cm$^{-3}$ and thereafter incoherently dedispersed to a range of nearby trial DMs with a step size of $0.0002$\,pc\,cm$^{-3}$. The peak S/N is determined for every DM and is shown with grey circles for B30 and grey pentagons for B43. For visual purposes, moving averages are shown with solid and dashed black lines for B30 and B43, respectively. A Lorentzian distribution is fit (solid green line) to the individual data points close to the peak of the profile of B30 and the best DM, $560.105$\,pc\,cm$^{-3}$, is determined to be the centre of the fitted distribution (solid vertical line). Similarly, all the data points of B43 are fitted with a Gaussian distribution (solid orange line), which peaks at a DM of $560.308$\,pc\,cm$^{-3}$ (dashed vertical line).}
    \label{fig:peaksn}
\end{figure*}

\clearpage

\begin{figure*}
    \centerline{\includegraphics[width=1.2\textwidth]{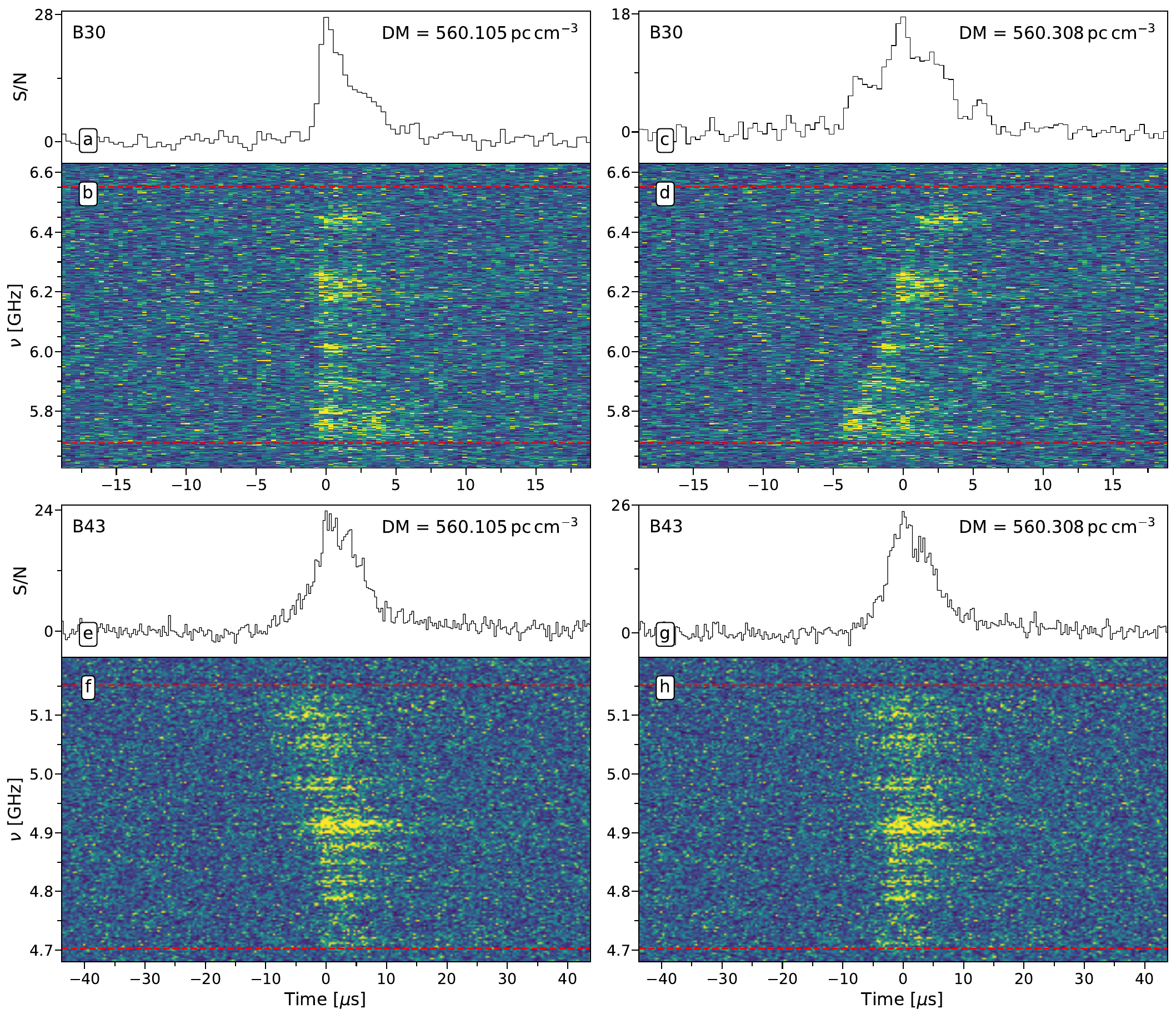}}
    \caption{\textbf{Dispersion measure (DM) comparison for the two brightest ultra-FRBs.} Burst B30 (panels a, b, c and d) and burst B43 (panels e, f, g and h) are coherently dedispersed to a DM of $560.105$\,pc\,cm$^{-3}$ (left column) and $560.308$\,pc\,cm$^{-3}$ (right column). The dynamic spectra (panels b, d, f and h)  have a time resolution of $341.\bar{3}$\,ns and a frequency resolution of $2.9296875$\,MHz. The dashed red lines indicate the frequency range that was averaged over to create the burst profiles, which also have a time resolution of $341.\bar{3}$\,ns (panels a, c, e and g). Using a DM of $560.308$\,pc\,cm$^{-3}$ slightly increases the peak S/N of burst B43 and the width of burst B43 decreases by about $2$\,$\upmu$s. However, using a DM of $560.308$\,pc\,cm$^{-3}$, burst B30 is clearly over-dedispersed (panel d).}
    \label{fig:b30b43_comp}
\end{figure*}

\clearpage

\begin{figure*}
    \centerline{\includegraphics[width=1.2\textwidth]{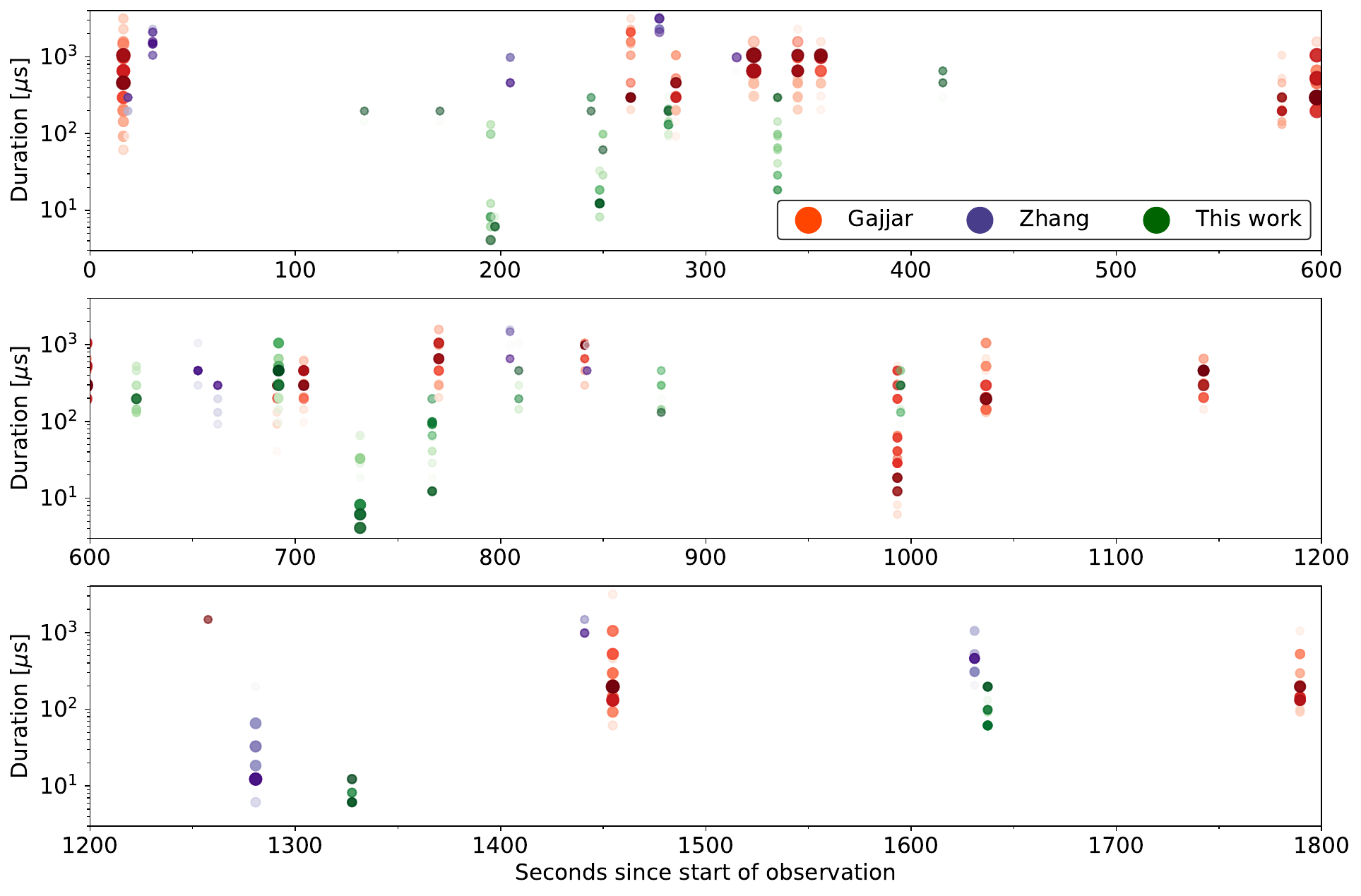}}
    \caption{\textbf{Detection duration as a function of time.} Every FRB is found in multiple subbands and/or trial boxcar widths (Methods). Both the colour shading and the size of the data points correspond to the S/N of the burst detection, with larger/darker dots indicating a higher S/N. Colours indicate whether the bursts have been found before by \citet{gajjar_2018_apj} and \citet{zhang_2018_apj}.}
    \label{fig:dur_vs_time}
\end{figure*}

\clearpage

\begin{figure*}
    \centerline{\includegraphics[width=1.2\textwidth]{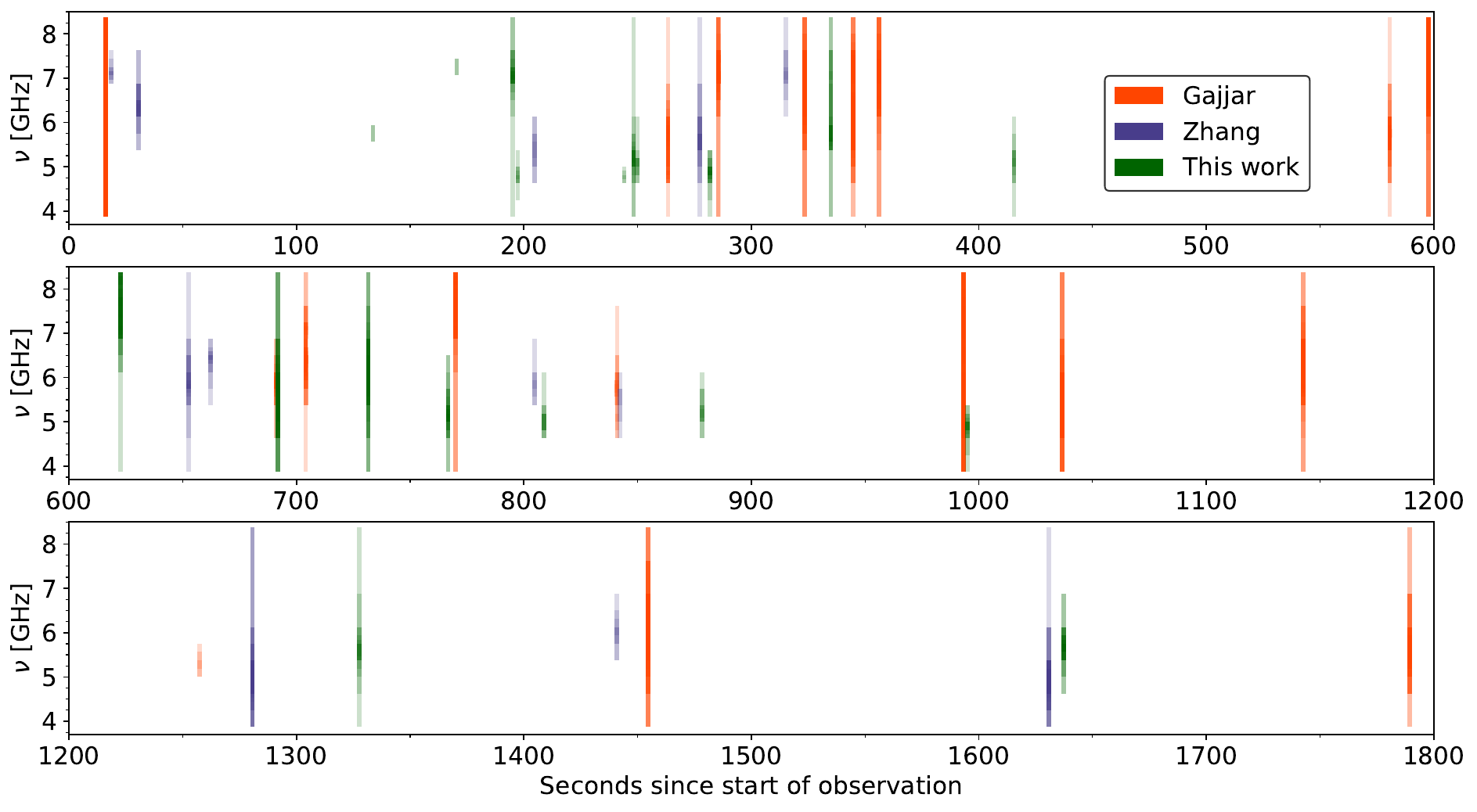}}
    \caption{\textbf{Spectral extent as a function of time.} Every FRB is found in multiple subbands and/or trial boxcar widths (Methods). For every burst detection we plot a rectangle showing the frequency range of the corresponding subband and a $2$-second time interval around the burst arrival time. Every rectangle has the same transparency and the colours become darker as multiple patches are plotted on top of each other. Colours indicate whether the bursts have been found before by \citet{gajjar_2018_apj} and \citet{zhang_2018_apj}.}
    \label{fig:spec_vs_time}
\end{figure*}

\clearpage

\begin{figure*}
    \centerline{\includegraphics[width=1.0\textwidth]{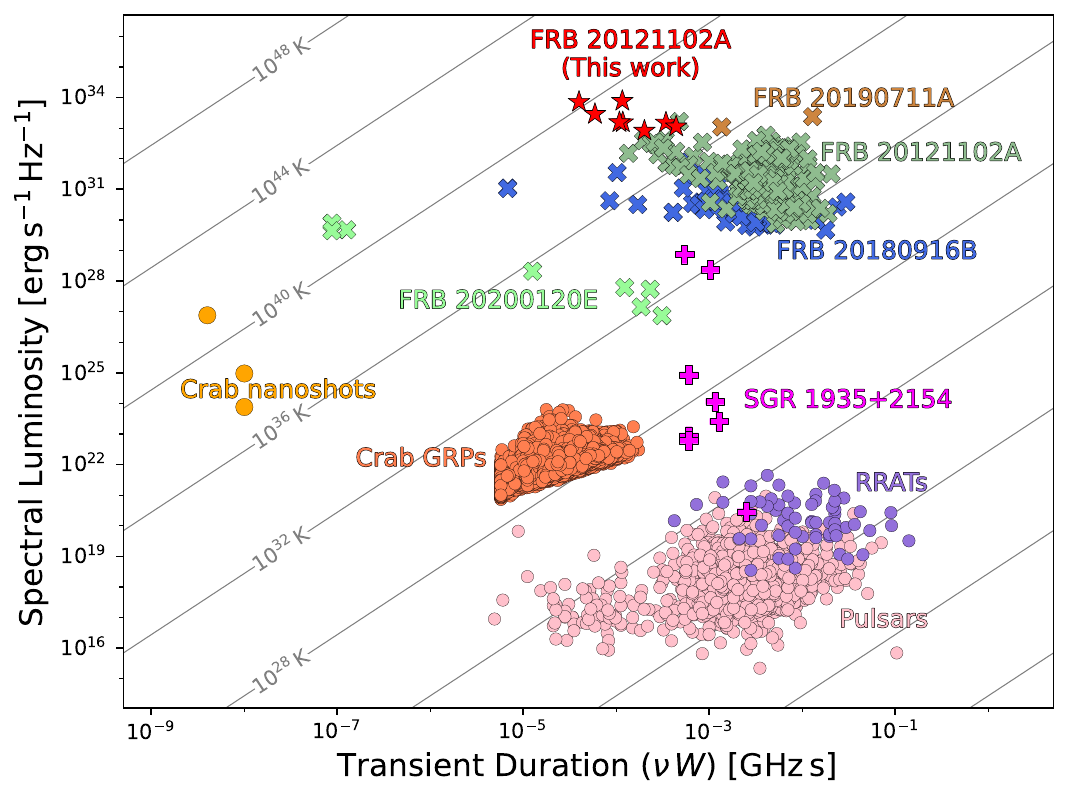}}
    \caption{\textbf{The transient phase space for coherent radio emission.} The transient duration (width times the central frequency of the burst) and the isotropic-equivalent spectral luminosity of the 8 ultra-FRBs shown in Figure~\ref{fig:family_plot} are plotted as red stars. A selection of published localised repeating FRBs are plotted as crosses. Radio bursts from the Galactic magnetar SGR~1935+2154 are shown as purple plusses. The pulsar and Rotating RAdio Transient (RRAT) population are shown as pink and magenta circles, respectively. Giant radio pulses (GRPs) and `nanoshots' of the Crab pulsar are shown as orange and yellow circles, respectively. The diagonal grey lines represent levels of constant brightness temperature. The three data points from FRB~20200120E that have a brightness temperature of $>10^{40}$\,K are from bursts that show short temporal structure within a broader burst envelope\cite{nimmo_2022_natas}. Based on a figure presented in \citet{nimmo_2022_natas} (see references therein).} 
    \label{fig:tps}
\end{figure*}

\clearpage

\begin{table*}
\caption{\textbf{Basic detection properties of the bursts.} The time of arrival (TOA), signal-to-noise (S/N), duration, probability and frequency extent are with respect to the detection with the highest S/N of that burst.}
\centering
\label{tab:burst_detec_prop}
\resizebox{\textwidth}{!}{%
\begin{tabular}{crrrrrcccc}
\hline
\hline
Burst & TOA$^{\alpha}$ & PRESTO & Duration     & Probability$^{\gamma}$ & Total      & $\nu_{\mathrm{low}}$$^{\delta}$ & $\nu_{\mathrm{high}}$$^{\delta}$ & \citet{gajjar_2018_apj} & \citet{zhang_2018_apj} \\
      & {[s]}        & S/N    & {[$\upmu$s]} & $\times 100\,\%$        & detections & {[MHz]}                       & {[MHz]}                                   & identifier                                 & identifier                                \\ \hline
B01   & 16.2701        & 74.05  & 458.752      & 100                    & 424        & 6970.00                         & 7157.50                          & 11A                                        & 1                                         \\
B02   & 18.4898        & 9.54   & 294.912      & 100                    & 6          & 6970.00                         & 7157.50                          &                                            & 2                                         \\
B03   & 30.5842        & 10.15  & 1572.864     & 100                    & 13         & 6126.25                         & 6501.25                          &                                            & 4                                         \\
B04   & 133.7138       & 8.36   & 196.608      & 100                    & 2          & 5563.75                         & 5938.75                          &                                            &                                           \\
B05   & 170.5020       & 8.31   & 196.608      & 100                    & 2          & 7063.75                         & 7438.75                          &                                            &                                           \\
B06   & 195.1935       & 14.11  & 4.096        & 100                    & 16         & 6970.00                         & 7157.50                          &                                            &                                           \\
B07   & 197.3795       & 10.58  & 6.144        & 0                      & 5          & 4720.00                         & 4907.50                          &                                            &                                           \\
B08   & 204.7666       & 9.53   & 458.752      & 100                    & 7          & 5001.25                         & 5751.25                          &                                            & 6                                         \\
B09   & 244.1764       & 8.82   & 196.608      & 100                    & 3          & 4626.25                         & 4813.75                          &                                            &                                           \\
B10   & 248.3161       & 12.47  & 12.288       & 0                      & 15         & 5001.25                         & 5751.25                          &                                            &                                           \\
B11   & 249.9378       & 8.84   & 61.440       & 100                    & 6          & 4813.75                         & 5188.75                          &                                            &                                           \\
B12   & 263.4282       & 16.01  & 294.912      & 100                    & 42         & 5376.25                         & 6126.25                          & 11B                                        & 8,9                                       \\
B13   & 277.4008       & 12.65  & 3145.728     & 100                    & 11         & 5376.25                         & 6126.25                          &                                            & 10                                        \\
B14   & 281.8991       & 12.47  & 196.608      & 100                    & 13         & 4813.75                         & 5001.25                          &                                            &                                           \\
B15   & 285.4547       & 23.44  & 458.752      & 100                    & 48         & 6970.00                         & 7157.50                          & 11C                                        & 11                                        \\
B16   & 315.0567       & 12.90  & 983.040      & 82                     & 5          & 6970.00                         & 7157.50                          &                                            & 12                                        \\
B17   & 323.3700       & 99.22  & 1048.576     & 95                     & 154        & 6876.25                         & 7251.25                          & 11D                                        & 13                                        \\
B18   & 334.9762       & 8.84   & 294.912      & 100                    & 20         & 5376.25                         & 6876.25                          &                                            &                                           \\
B19   & 344.7870       & 41.62  & 1048.576     & 69                     & 105        & 6970.00                         & 7157.50                          & 11E                                        & 14                                        \\
B20   & 356.0529       & 55.87  & 1048.576     & 50                     & 47         & 6970.00                         & 7157.50                          & 11F                                        & 15                                        \\
B21   & 415.5190       & 8.76   & 458.752      & 100                    & 6          & 4626.25                         & 5376.25                          &                                            &                                           \\
B22   & 580.6662       & 14.18  & 294.912      & 100                    & 34         & 5376.25                         & 6126.25                          & 11G                                        & 19                                        \\
B23   & 597.6259       & 95.94  & 294.912      & 100                    & 128        & 6876.25                         & 7626.25                          & 11H                                        & 20                                        \\
B24   & 622.5689       & 17.49  & 196.608      & 100                    & 42         & 6876.25                         & 8376.25                          &                                            &                                           \\
B25   & 652.5810       & 9.52   & 458.752      & 100                    & 12         & 5376.25                         & 6876.25                          &                                            & 21                                        \\
B26   & 662.2081       & 9.45   & 294.912      & 100                    & 8          & 5751.25                         & 6501.25                          &                                            & 22                                        \\
B27   & 691.0620       & 10.96  & 294.912      & 100                    & 18         & 5751.25                         & 6501.25                          & 11I                                        & 23                                        \\
B28   & 691.8646       & 25.21  & 458.752      & 100                    & 77         & 5376.25                         & 6126.25                          &                                            &                                           \\
B29   & 704.0925       & 21.31  & 294.912      & 100                    & 49         & 6126.25                         & 6501.25                          & 11J                                        & 24                                        \\
B30   & 731.5633       & 33.13  & 4.096        & 20                     & 56         & 5751.25                         & 6501.25                          &                                            &                                           \\
B31   & 766.6877       & 11.15  & 12.288       & 5                      & 25         & 5001.25                         & 5751.25                          &                                            &                                           \\
B32   & 769.8707       & 21.22  & 655.360      & 100                    & 51         & 6876.25                         & 8376.25                          & 11K                                        & 25                                        \\
B33   & 804.5907       & 8.24   & 655.360      & 100                    & 4          & 5563.75                         & 5938.75                          &                                            & 26                                        \\
B34   & 808.8735       & 8.93   & 458.752      & 100                    & 7          & 4813.75                         & 5188.75                          &                                            &                                           \\
B35   & 840.9772       & 9.55   & 983.040      & 100                    & 14         & 5563.75                         & 5938.75                          & 11L                                        & 27,28,29                                  \\
B36   & 842.1029       & 8.52   & 458.752      & 100                    & 3          & 5001.25                         & 5751.25                          &                                            & 31                                        \\
B37   & 878.2482       & 8.80   & 131.072      & 100                    & 6          & 5001.25                         & 5751.25                          &                                            &                                           \\
B38   & 993.2905       & 14.50  & 12.288       & 100                    & 77         & 5657.50                         & 5845.00                          & 11M                                        & 33                                        \\
B39   & 994.9033       & 10.35  & 294.912      & 100                    & 11         & 4813.75                         & 5188.75                          &                                            &                                           \\
B40   & 1036.4686      & 35.74  & 196.608      & 100                    & 102        & 5001.25                         & 5751.25                          & 11N                                        & 34                                        \\
B41   & 1142.4320      & 28.03  & 458.752      & 100                    & 71         & 5376.25                         & 6876.25                          & 11O                                        & 37                                        \\
B42   & 1257.4688      & 8.32   & 1474.560     & 100                    & 3          & 5001.25                         & 5376.25                          & 11P                                        & 39                                        \\
B43   & 1280.6520      & 46.14  & 12.288       & 100                    & 47         & 4813.75                         & 5188.75                          &                                            & 40                                        \\
B44   & 1327.5579      & 12.91  & 12.288       & 0                      & 12         & 5001.25                         & 5751.25                          &                                            &                                           \\
B45   & 1440.8676      & 10.10  & 983.040      & 100                    & 5          & 5751.25                         & 6501.25                          &                                            & 42                                        \\
B46   & 1454.5774      & 55.57  & 196.608      & 100                    & 129        & 5751.25                         & 6501.25                          & 11Q                                        & 43                                        \\
B47   & 1630.8575      & 19.32  & 458.752      & 100                    & 46         & 4626.25                         & 5001.25                          &                                            & 45                                        \\
B48   & 1637.2393      & 12.24  & 196.608      & 100                    & 22         & 5657.50                         & 5845.00                          &                                            &                                           \\
B49   & 1789.3961      & 30.26  & 196.608      & 100                    & 55         & 5376.25                         & 5751.25                          & 11R                                        & 46                                        \\ \hline

\multicolumn{10}{l}{$^{\alpha}$ Seconds since the start of the observation (MJD~$57991.57760417$).} \\

\multicolumn{10}{l}{$^{\gamma}$ The \texttt{FETCH}\cite{agarwal_2020_mnras} probability of the candidate being of astrophysical origin (model A)} \\
\multicolumn{10}{l}{\hspace{0.25cm} indicating the importance of manually checking the candidates with duration $\leq500$\,$\upmu$s.} \\ 

\multicolumn{10}{l}{$^{\delta}$ For clarity $0.214844$\,MHz has been subtracted from every number.} \\

\hline
\end{tabular}
}
\end{table*}

\clearpage

\begin{table*}
\caption{\textbf{Frequency coverage per compute node.}}
\centering
\label{tab:nodes}
\begin{tabular}{lcccc} 
\hline
\hline
Node &
  \begin{tabular}[c]{@{}c@{}}Passband$^{\gamma}$\end{tabular} &
  \begin{tabular}[c]{@{}c@{}}$\nu_\textrm{low}^{\alpha}$\\ {[}MHz{]}\end{tabular} &
  \begin{tabular}[c]{@{}c@{}}$\nu_\textrm{center}^{\alpha}$\\ {[}MHz{]}\end{tabular} &
  \begin{tabular}[c]{@{}c@{}}$\nu_\textrm{high}^{\alpha}$\\ {[}MHz{]}\end{tabular} \\ \hline
BLP00                      & 0 & 9126.25 & 9220.00 & 9313.75 \\
BLP01                      & 0 & 8938.75 & 9032.50 & 9126.25 \\
BLP02                      & 0 & 8751.25 & 8845.00 & 8938.75 \\
BLP03                      & 0 & 8563.75 & 8657.50 & 8751.25 \\
BLP04                      & 0 & 8376.25 & 8470.00 & 8563.75 \\
BLP05$^{\dagger}$          & 0 & 8188.75 & 8282.50 & 8376.25 \\
BLP06$^{\dagger}$          & 0 & 8001.25 & 8095.00 & 8188.75 \\
BLP07$^{\triangle}$            & 0 & 7813.75 & 7907.50 & 8001.25 \\
BLP10$^{\dagger\triangle}$     & 1 & 7813.75 & 7907.50 & 8001.25 \\
BLP11$^{\dagger}$          & 1 & 7626.25 & 7720.00 & 7813.75 \\
BLP12$^{\dagger}$          & 1 & 7438.75 & 7532.50 & 7626.25 \\
BLP13$^{\dagger}$          & 1 & 7251.25 & 7345.00 & 7438.75 \\
BLP14$^{\dagger}$          & 1 & 7063.75 & 7157.50 & 7251.25 \\
BLP15$^{\dagger}$          & 1 & 6876.25 & 6970.00 & 7063.75 \\
BLP16$^{\dagger}$          & 1 & 6688.75 & 6782.50 & 6876.25 \\
BLP17$^{\triangle}$        & 1 & 6501.25 & 6595.00 & 6688.75 \\
BLP20$^{\dagger\triangle}$ & 2 & 6501.25 & 6595.00 & 6688.75 \\
BLP21$^{\dagger}$          & 2 & 6313.75 & 6407.50 & 6501.25 \\
BLP22$^{\dagger}$          & 2 & 6126.25 & 6220.00 & 6313.75 \\
BLP23$^{\dagger}$          & 2 & 5938.75 & 6032.50 & 6126.25 \\
BLP24$^{\dagger}$          & 2 & 5751.25 & 5845.00 & 5938.75 \\
BLP25$^{\dagger}$          & 2 & 5563.75 & 5657.50 & 5751.25 \\
BLP26$^{\dagger}$          & 2 & 5376.25 & 5470.00 & 5563.75 \\
BLP27$^{\triangle}$          & 2 & 5188.75 & 5282.50 & 5376.25 \\
BLP30$^{\dagger\triangle}$   & 3 & 5188.75 & 5282.50 & 5376.25 \\
BLP31$^{\dagger}$          & 3 & 5001.25 & 5095.00 & 5188.75 \\
BLP32$^{\dagger}$          & 3 & 4813.75 & 4907.50 & 5001.25 \\
BLP33$^{\dagger}$          & 3 & 4626.25 & 4720.00 & 4813.75 \\
BLP34$^{\dagger}$          & 3 & 4438.75 & 4532.50 & 4626.25 \\
BLP35$^{\dagger}$          & 3 & 4251.25 & 4345.00 & 4438.75 \\
BLP36$^{\dagger}$          & 3 & 4063.75 & 4157.50 & 4251.25 \\
BLP37$^{\dagger}$          & 3 & 3876.25 & 3970.00 & 4063.75 \\ \hline

\multicolumn{5}{l}{$^{\gamma}$ The passbands were numbered 0--3 and were tuned to a} \\
\multicolumn{5}{l}{\hspace{0.25cm} centre frequency of $8563.964844$, $7251.464844$,} \\ 
\multicolumn{5}{l}{\hspace{0.25cm} $5938.964844$ and $4626.464844$\,MHz, respectively.} \\ 

\multicolumn{5}{l}{$^{\alpha}$ For clarity $0.214844$\,MHz has been subtracted from} \\
\multicolumn{5}{l}{\hspace{0.25cm} every value.} \\

\multicolumn{5}{l}{$^{\dagger}$ The nodes that were used to create high-time-resolution} \\
\multicolumn{5}{l}{\hspace{0.25cm} filterbanks for the burst search.} \\

\multicolumn{5}{l}{$^{\triangle}$ These nodes have exactly $187.5$\,MHz overlap} \\
\multicolumn{5}{l}{\hspace{0.55cm}with an adjacent node of a different passband.} \\ 

\hline
\end{tabular}
\end{table*}

\clearpage

\begin{table*}
    \centering
    \caption{\textbf{Search strategy.} Dispersion measure (DM) range, DM step-size ($\Delta$DM) and number of `subbands' used in \texttt{PRESTO}'s \texttt{prepsubband} as a function of sampling time ($\Delta t$), bandwidth (BW) and frequency ($\nu$) range. For clarity we have subtracted $0.214844$\,MHz from all the shown frequencies. Full table available online}.\label{tab:prepsubband}
    \begin{tabular}{ccccccccccc}
    \hline \hline
    $\Delta t$ & BW & Channels & $\nu_{\mathrm{low}}$ & $\nu_{\mathrm{center}}$ & $\nu_{\mathrm{high}}$ & DM$_{\mathrm{low}}$ & DM$_{\mathrm{high}}$ & Trial DMs & $\Delta$DM & Subbands \\
    $\left[ \upmu \mathrm{s} \right]$ & [MHz] & N & [MHz] & [MHz] & [MHz] & [pc cm$^{-3}$] & [pc cm$^{-3}$] & N & [pc cm$^{-3}$] & N \\
    \hline
2.048 & 4500.0 & 1536 & 3876.25 & 6126.25 & 8376.25 & 552.90 & 566.90 & 701 & 0.02 & 128 \\
2.048 & 1500.0 & 512 & 3876.25 & 4626.25 & 5376.25 & 552.91 & 566.89 & 467 & 0.03 & 64 \\
2.048 & 1500.0 & 512 & 4626.25 & 5376.25 & 6126.25 & 552.90 & 566.90 & 281 & 0.05 & 64 \\
2.048 & 1500.0 & 512 & 5376.25 & 6126.25 & 6876.25 & 552.90 & 566.90 & 281 & 0.05 & 64 \\
2.048 & 1500.0 & 512 & 6126.25 & 6876.25 & 7626.25 & 552.90 & 566.90 & 141 & 0.10 & 64 \\
2.048 & 1500.0 & 512 & 6876.25 & 7626.25 & 8376.25 & 552.90 & 566.90 & 141 & 0.10 & 64 \\
2.048 & 750.0 & 256 & 3876.25 & 4251.25 & 4626.25 & 552.90 & 566.90 & 281 & 0.05 & 32 \\
2.048 & 750.0 & 256 & 4251.25 & 4626.25 & 5001.25 & 552.90 & 566.90 & 281 & 0.05 & 32 \\
\multicolumn{1}{c}{\vdots} & \multicolumn{1}{c}{\vdots} & \multicolumn{1}{c}{\vdots} & \multicolumn{1}{c}{\vdots} & \multicolumn{1}{c}{\vdots} & \multicolumn{1}{c}{\vdots} & \multicolumn{1}{c}{\vdots} & \multicolumn{1}{c}{\vdots} & \multicolumn{1}{c}{\vdots} & \multicolumn{1}{c}{\vdots} & \multicolumn{1}{c}{\vdots} \\
\hline
\end{tabular}
\end{table*}

\end{document}